\documentclass[12pt]{article}

\usepackage{epsfig}
\usepackage{epsfig,amsmath} 
\usepackage{graphics}

\newcommand{\be}{\begin{equation}}
\newcommand{\ee}{\end{equation}}
\newcommand{\bi}[1]{\vspace{-3mm} \bibitem{#1}}

\begin{document}

\today

\begin{center}
{\Large \bf 
Fractional dynamics of coupled oscillators with long-range interaction }
\vskip 5 mm

{\large \bf Vasily E. Tarasov$^{1,2}$ and George M. Zaslavsky$^{2,3}$ } \\

\vskip 3mm

{\it $1)$ Skobeltsyn Institute of Nuclear Physics, \\
Moscow State University, Moscow 119992, Russia } \\
{\it $2)$ Courant Institute of Mathematical Sciences, New York University \\
251 Mercer Street, New York, NY 10012, USA, }\\ 
{\it $3)$ Department of Physics, New York University, \\
2-4 Washington Place, New York, NY 10003, USA } \\
\end{center}

\vskip 11 mm

\begin{abstract}
We consider one-dimensional chain of coupled linear and nonlinear 
oscillators with long-range power-wise interaction.
The corresponding term in dynamical equations is proportional 
to $1/|n-m|^{\alpha+1}$. 
It is shown that the equation of motion in the infrared limit can 
be transformed into the medium equation
with the Riesz fractional derivative of order $\alpha$, 
when $0<\alpha<2$.
We consider few models of coupled oscillators and show 
how their synchronization can appear as a result of bifurcation,
and how the corresponding solutions depend on $\alpha$.
The presence of fractional derivative leads also to 
the occurrence of localized  structures.
Particular solutions for fractional time-dependent complex
Ginzburg-Landau (or nonlinear Schrodinger) equation are derived.
These solutions are interpreted as 
synchronized states and localized structures 
of the oscillatory medium.
\end{abstract}

\vskip 3 mm
{\small 



\vskip 3 mm


\vskip 11 mm

{\bf 
Although the fractional calculus is known for more than two hundred years
and its developing is an active area of mathematics,
appearance and use of it in physical literature
is fairly recent and sometimes is considered as exotic.
In fact, there are many different areas where fractional equations,
i.e., equations with fractional integro-differentiation,
describe real processes.
Between the most related areas are chaotic dynamics
\cite{Zaslavsky1}, random walk in fractal space-time \cite{Montr}
and random processes of Levy type \cite{SZ,Uch,MS1,MS2}.
The physical reasons for the appearance of fractional equations
are intermittancy, dissipation, wave propagation in complex media,
long memory, and others. This article deals with long-range 
interaction that can work in some way as a long memory.
A unified approach to the origin of fractional dynamics
from the long-range interaction of nonlinear oscillators 
or other objects permits to consider such phenomena as 
synchronization, breathers formation, space-time structures 
by the same formalism using new tools from the fractional calculus.
}

\section{Introduction}

Collective oscillation and synchronization are the fundamental 
phenomena in physics, chemistry, biology, and neuroscience,
which are actively studied recently \cite{Afr,Pik1,BKOVZ}, 
having both important theoretical and applied significance.
Beginning with the pioneering contributions 
by Winfree \cite{Win} and Kuramoto \cite{Kur1}, 
studies of synchronization in populations of coupled oscillators
becomes an active field of research in biology and chemistry.
An oscillatory medium is an extended system, where each site (element)
performs self-sustained oscillations.
A good physical and chemical example is the oscillatory Belousov-Zhabotinsky reaction
\cite{Belousov,Zhab,Kur1} in a medium where different 
sites can oscillate with different periods and phases.
Typically, the reaction is accompanied by a color variation of the medium. 
Complex Ginzburg-Landau equation is canonical model for oscillatory 
systems with local coupling near Hopf bifurcation.
Recently, Tanaka and Kuramoto \cite{TK} have shown how, in the vicinity 
of the bifurcation, the description of an array of nonlocally coupled 
oscillators can be reduced to the complex Ginzburg-Landau equation.
In Ref. \cite{Mikh}, a model of population of diffusively coupled 
oscillators with limit-cycles is described by the complex Ginzburg-Landau
equation with nonlocal interaction.
Nonlocal coupling is considered in Refs. \cite{Kur3,Kur4,Mikh}. 
The long-range interaction that decreases as $1/|x|^{\alpha+1}$ 
with $0< \alpha <2$ is considered in Refs. 
\cite{Dyson,J,CMP,NakTak,S} with respect to the system's
thermodynamics and phase transition.
It is also shown in \cite{Lask} that using the Fourier transform and 
limit for the wave number $k \rightarrow 0$, the long-range term interaction
leads under special conditions to the fractional dynamics.

In the last decade it is found that many physical processes can be adequately 
described by equations that consist of derivatives of fractional order.
In a fairly short period of time the list of such 
applications becomes long and the area of applications is broad.
Even in a concise form, the applications include material
sciences \cite{Hilfer,C2,Nig1,Nig3}, chaotic dynamics \cite{Zaslavsky1},
quantum theory \cite{Laskin,Naber,Krisch,Goldfain}, 
physical kinetics \cite{Zaslavsky1,SZ,ZE,Zaslavsky7},
fluids and plasma physics \cite{CLZ,Plasma2005}, 
and many others physical topics
related to wave propagation \cite{ZL}, long-range dissipation \cite{GM}, 
anomalous diffusion and transport theory 
(see reviews \cite{Zaslavsky1,Montr,Hilfer,Uch,MK}).

It is known that the appearance of fractional derivatives 
in equations of motion can be linked to nonlocal properties of dynamics.
Fractional Ginzburg-Landau equation has been suggested in
\cite{Zaslavsky6,Physica2005,Mil}.
In this paper, we consider the synchronization for oscillators with
long-range interaction that in continuous limit leads to the
fractional complex Ginzburg-Landau equation. 
We confirm the result obtained in \cite{Lask} that
the infrared limit (wave number $k \rightarrow 0$)
of an infinite chain of oscillators with the long-range interaction
can be described by equations with fractional Riesz 
coordinate derivative of order $\alpha<2$.
This result permits to apply different tools of the fractional calculus
to the considered systems, and to interpret different system's 
features in an unified way.

In Sec. 2, 
we consider a systems of oscillators with linear long-range interaction. 
For infrared behavior of the oscillatory medium, 
we obtain the equations that has coordinate derivatives
of fractional order. 
In Sec. 3, some particular solutions are derived with
a constant wave number for the fractional Ginzburg-Landau equation. 
These solutions are interpreted as 
synchronization in the oscillatory medium.
In Sec. 4, we derive solutions of the fractional 
Ginzburg-Landau equation  near a limit cycle. 
These solutions are interpreted as coherent structures in 
the oscillatory medium with long-range interaction.
In Sec. 5, we consider the nonlinear long-range interaction of 
oscillators and corresponding equations for the spin field. 
Finally, discussion of the results and conclusion are given in Sec. 6.
In two appendices there are details of derivation of 
fractional phase (spin) equation.

\section{Long-range interaction of oscillators}

\subsection{Linear nonlocal coupling}

Let us consider the oscillators that are described by 
\be \label{E1}
\frac{d}{dt}z_n(t)=F(z_n) ,
\ee
where $z_n$ is the position of the $n$-th oscillator in the complex plane,
and $F$ is a force.
As an example, for the oscillators with a limit-cycle, $F$ 
can be taken as
\be \label{E2}
F(z)=(1+ia)z-(1+ib) |z|^2 z .
\ee
For $a=b$, each oscillator has a stable limit cycle at $|z_n|=1$.
Consider an infinite population ($N \rightarrow \infty$) of oscillators (\ref{E1}),
with linear nonlocal coupling 
\be \label{C1}
\frac{d}{dt}z_n(t)=F(z_n)+g_0\sum_{m \not=n} J_{\alpha}(n-m) (z_n-z_m) ,
\ee
where the nonlocality is given by the power function
\be \label{C2}
J_{\alpha}(n)= |n|^{-\alpha-1} .
\ee
This coupling in the limit $\alpha \rightarrow \infty$ 
is a nearest-neighbor interaction.

\subsection{Derivation of equation for continuous oscillatory medium}

In this section, we derive the equations for continuous medium 
that consists of oscillators (\ref{C1}) with long-range interaction (\ref{C2}).
Let us define field
\be \label{Zxt}
Z(x,t)=\frac{1}{2 \pi} \int^{+\infty}_{-\infty} dk \ 
e^{ikx} \sum^{+\infty}_{n=-\infty} e^{-ikn} z_n(t) 
\ee
to describe the continuous oscillatory medium.

Multiplying Eq. (\ref{C1}) on $\exp(-ikn)$, 
and summing over $n$ from $-\infty$ to $+\infty$, we obtain
\be \label{C3}
\sum^{\infty}_{n=-\infty} e^{-ikn} \frac{d}{dt}z_n(t)=\sum^{\infty}_{n=-\infty} e^{-ikn} F(z_n)+
g_0 \sum^{\infty}_{n=-\infty} \sum_{m \not=n}e^{-ikn} \frac{1}{|n-m|^{\alpha+1}} (z_n-z_m) .
\ee
It is convenient to introduce a new field variable.  
Let us define
\be \label{C4}
y(k,t)=\sum^{\infty}_{n=-\infty} e^{-ikn} z_n(t),
\ee
\be \label{C5}
\tilde J_{\alpha}(k)=\sum_{n\not=0} e^{-ikn} J_{\alpha}(n)=
\sum_{n\not=0} e^{-ikn} \frac{1}{|n|^{\alpha+1}} .
\ee
The interaction term of (\ref{C3}) can be presented as two terms
\[
\sum^{\infty}_{n=-\infty} \sum_{m \not=n} e^{-ikn} \frac{1}{|n-m|^{\alpha+1}} (z_n-z_m) = \]
\be \label{C6}
=\sum^{\infty}_{n=-\infty} \sum_{m \not=n} e^{-ikn} \frac{1}{|n-m|^{\alpha+1}} z_n - 
\sum^{\infty}_{n=-\infty} \sum_{m \not=n} e^{-ikn} \frac{1}{|n-m|^{\alpha+1}} z_m .
\ee
For the first term in r.h.s. of (\ref{C6}):
\[ 
\sum^{\infty}_{n=-\infty} \sum_{m \not=n} e^{-ikn} \frac{1}{|n-m|^{\alpha+1}} z_n =
\sum^{\infty}_{n=-\infty} e^{-ikn} z_n \sum_{m \not=n}\frac{1}{|n-m|^{\alpha+1}} =
\]
\be \label{C7}
=\sum^{\infty}_{n=-\infty} e^{-ikn} z_n \sum_{m^{\prime} \not=0} \frac{1}{|m^{\prime}|^{\alpha+1}} =
y(k,t) \tilde J_{\alpha}(0) ,
\ee
where
\be \label{C8}
\tilde J_{\alpha}(0)=\sum_{n \not=0} \frac{1}{|n|^{\alpha+1}}=
2\sum^{\infty}_{n=1} \frac{1}{|n|^{\alpha+1}}= 2 \zeta(\alpha+1) ,
\ee
and $\zeta(z)$ is the Riemann zeta-function.
For the second term in r.h.s. of (\ref{C6}):
\[
\sum^{\infty}_{n=-\infty} \sum_{m \not=n} e^{-ikn} \frac{1}{|n-m|^{\alpha+1}} z_m = 
\sum^{\infty}_{m=-\infty} z_m \sum_{n\not=m} e^{-ikn} \frac{1}{|n-m|^{\alpha+1}} = \]
\be \label{C9}
=\sum_{m } z_m e^{-ikm}
\sum_{n^{\prime}\not=0} e^{-ikn^{\prime}} \frac{1}{|n^{\prime}|^{\alpha+1}} = 
y(k,t)\tilde J_{\alpha}(k) ,
\ee
where $\tilde J_{\alpha}(k)$ is defined by (\ref{C5}). 

As the result, Eq. (\ref{C3}) is
\be \label{C10}
\frac{\partial}{\partial t}y(k,t)=\hat F(y(k,t))+
g_0[\tilde J_{\alpha}(0)- \tilde J_{\alpha}(k)] y(k,t) ,
\ee 
where $\hat F(y(k,t))$ is an operator notation for the Fourier
transform of $F(z_n)$.
The function $\tilde J_{\alpha}(k)$ introduced in (\ref{C5}) 
can be transformed as
\[
\tilde J_{\alpha}(k)=
\sum_{n\not=0} e^{-ikn} \frac{1}{|n|^{\alpha+1}}=
\sum^{\infty}_{n=1} e^{-ikn} \frac{1}{|n|^{\alpha+1}}
+\sum^{-\infty}_{n=-1} e^{-ikn} \frac{1}{|n|^{\alpha+1}}=
\]
\[
=\sum^{\infty}_{n=1} e^{-ikn} \frac{1}{|n|^{\alpha+1}}
+\sum^{\infty}_{n=1} e^{ikn} \frac{1}{|n|^{\alpha+1}}=
\sum^{\infty}_{n=1} \frac{1}{n^{\alpha+1}} \left( e^{-ikn} +e^{ikn} \right)=
\]
\be \label{C11}
=Li_{\alpha+1}( e^{ik} ) +Li_{\alpha+1}( e^{-ik} ) ,
\ee
where $Li_{\alpha}(z)$ is a polylogarithm function.
This presentation was also obtained in \cite{Lask},
and it plays an important role in the following transition 
to fractional dynamics of $\tilde J_{\alpha}(k)$ through the polylogarithm.
Using the expansion
\be \label{D1}
Li_{\beta}(e^z)=\Gamma(1-\beta) (-z)^{\beta-1}+\sum^{\infty}_{n=0}
\frac{\zeta(\beta-n)}{n!} z^n, \quad |z|< 2\pi ,
\ee
we obtain
\be \label{D2}
\tilde J_{\alpha}(k)=2\Gamma(-\alpha)\cos(\pi \alpha/2) |k|^{\alpha}+
2\sum^{\infty}_{n=0} \frac{\zeta(\alpha+1-2n)}{(2n)!} (-k^2)^n , \quad
\tilde J_{\alpha}(0)=2 \zeta(\alpha+1) .
\ee
Using (\ref{C11}) in the form
\be
\tilde J_{\alpha}(k)=2\sum^{\infty}_{n=1} \frac{\cos(kn)}{n^{\alpha+1}}, 
\ee
we can see that
\be
\tilde J_{\alpha}(k+2\pi m)=\tilde J_{\alpha}(k) ,
\ee
where $m$ is an integer. 
For $\alpha=2$, $J_{\alpha}(k)$ is the Clausen function $Cl_{3}(k)$ \cite{Le}.
The plots of $J_{\alpha}(k)$ for $\alpha=1.1$, and $\alpha=1.9$
are presented in Fig. 1.


\begin{figure}
\centering
\rotatebox{270}{\includegraphics[width=7 cm,height=7 cm]{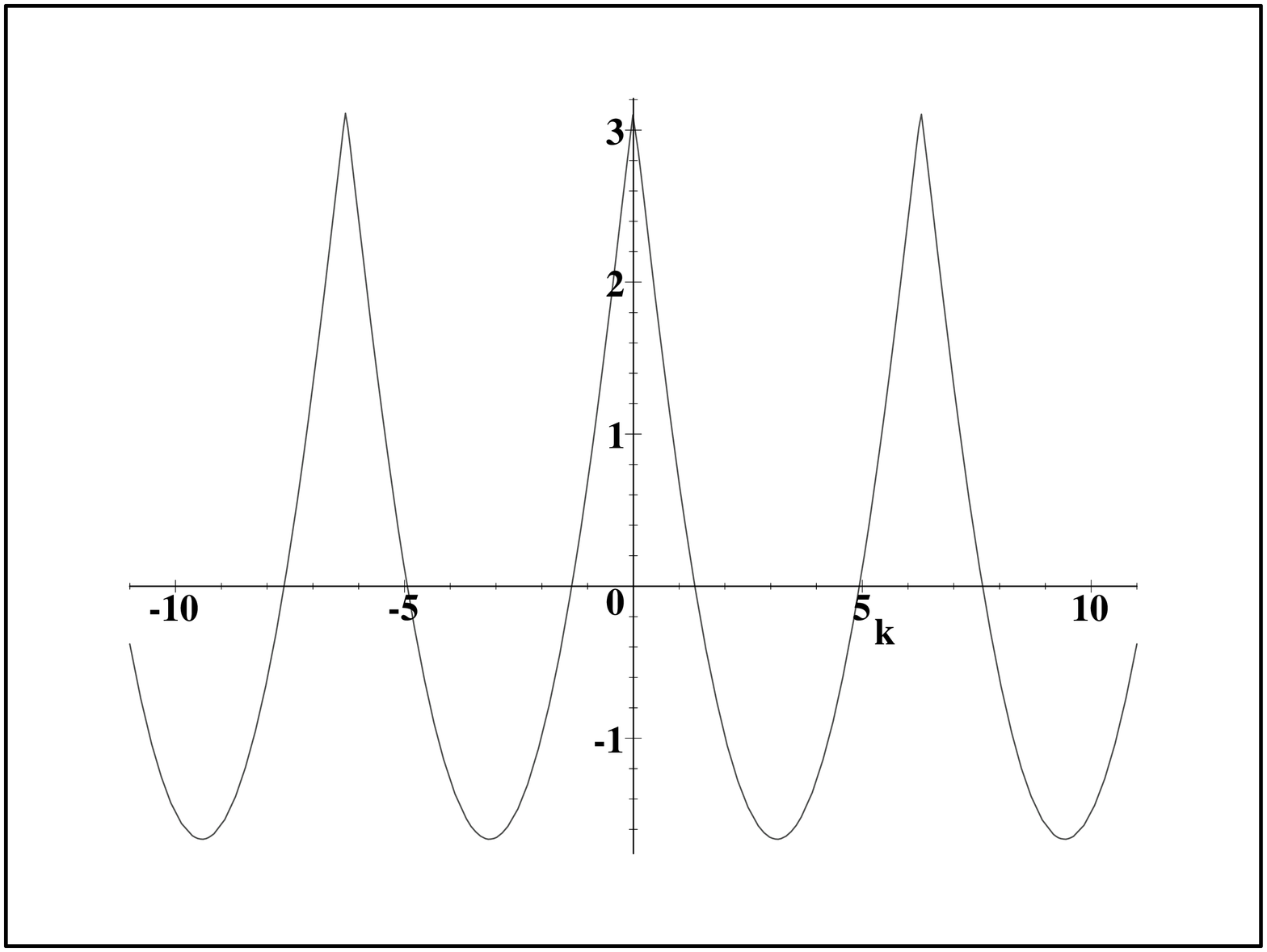}}
\rotatebox{270}{\includegraphics[width=7 cm,height=7 cm]{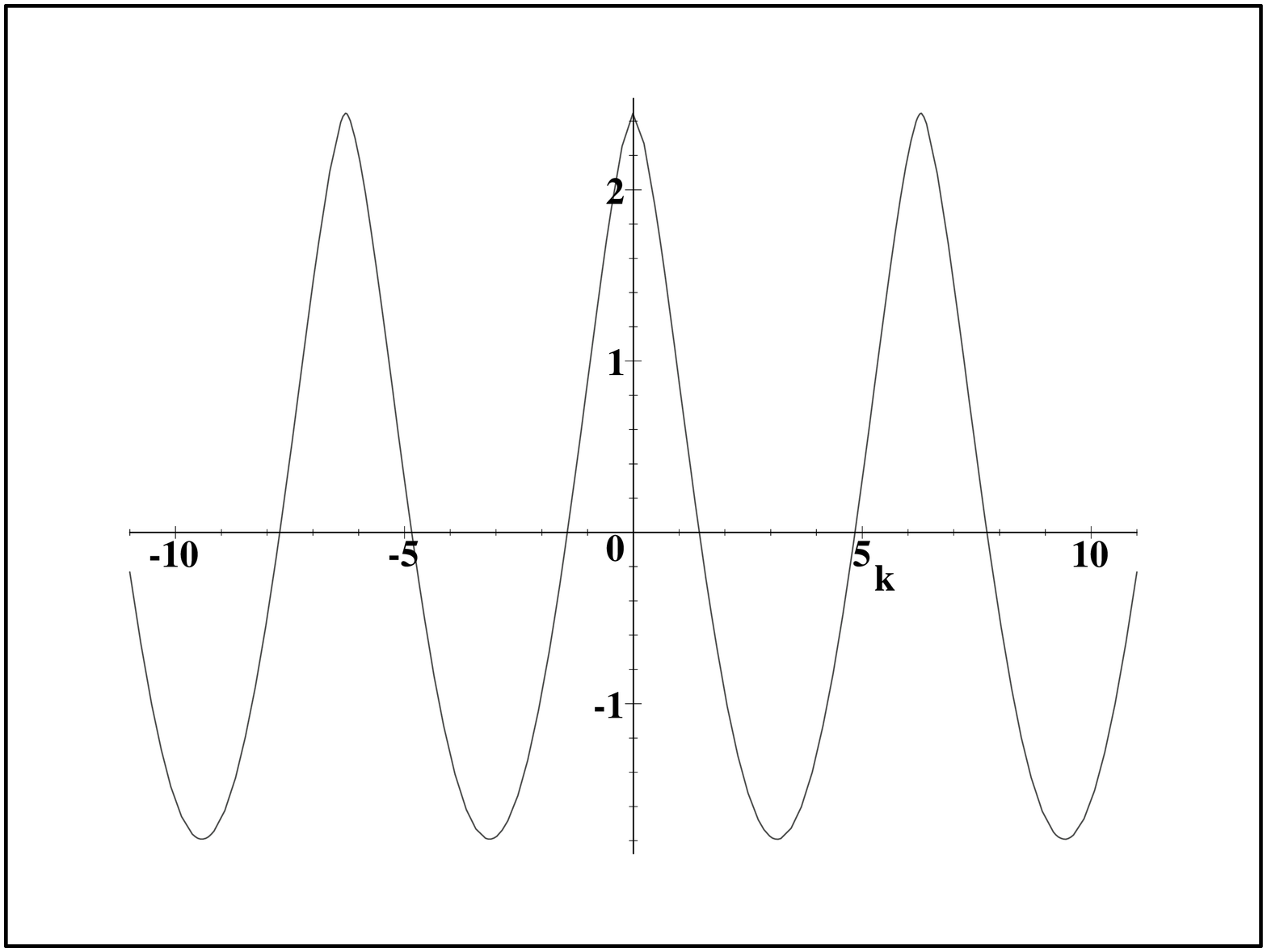}}
\caption{\label{fig1J} 
The function $J_{\alpha}(k)$ for orders $\alpha=1.1$, and $\alpha=1.9$.
}
\end{figure}

Then Eq. (\ref{C10}) has the form
\be \label{D4}
\frac{\partial}{\partial t}y(k,t)=\hat F(y(k,t))-
g_0 a_{\alpha} |k|^{\alpha} y(k,t) -
2g_0 \sum^{\infty}_{n=1} \frac{\zeta(\alpha+1-2n)}{(2n)!} (-k^2)^n y(k,t),
\ee
where 
\be \label{D5}
a_{\alpha} =2\Gamma(-\alpha)\cos(\pi \alpha/2) , \quad (0< \alpha <2, \ \alpha \not=1) .
\ee

To derive the equation for field (\ref{Zxt}), 
we can use definition (\ref{Zxt})
\be \label{D6}
Z(x,t)=\frac{1}{2\pi} \int^{\infty}_{-\infty} e^{ikx} y(k,t) dk,
\ee
and the connection between Riesz fractional derivative 
and its Fourier transform \cite{SKM}: 
\be
|k|^{\alpha} \longleftrightarrow
- \frac{\partial^{\alpha}}{\partial |x|^{\alpha}} ,
\quad
k^2 \longleftrightarrow
- \frac{\partial^2}{\partial |x|^2} .
\ee
The properties of the Riesz derivative can be found in 
\cite{SKM,OS,MR,Podlubny}. Its another expression is
\be \label{R1}
\frac{\partial^{\alpha}}{\partial |x|^{\alpha}} Z(x,t)=
-\frac{1}{2 \cos(\pi \alpha /2)} 
\left({\cal D}^{\alpha}_{+}Z(x,t) +{\cal D}^{\alpha}_{-} Z(x,t)\right) ,
\ee
where $\alpha\not=1,3,5...$, and
${\cal D}^{\alpha}_{\pm}$ are Riemann-Liouville 
left and right fractional derivatives
\[
{\cal D}^{\alpha}_{+}Z(x,t)=
\frac{1}{\Gamma(n-\alpha)} \frac{\partial^n}{\partial x^n}
\int^{x}_{-\infty} \frac{Z(\xi,t) d\xi}{(x-\xi)^{\alpha-n+1}},
\]
\be \label{R3}
{\cal D}^{\alpha}_{-}Z(x,t)=
\frac{(-1)^n}{\Gamma(n-\alpha)} \frac{\partial^n}{\partial x^n}
\int^{\infty}_x \frac{Z(\xi,t) d\xi}{(\xi-x)^{\alpha-n+1}} .
\ee
Substitution of Eqs. (\ref{R3}) into Eq. (\ref{R1}) gives
\be \label{R4}
\frac{\partial^{\alpha}}{\partial |x|^{\alpha}} Z(x,t)=
\frac{-1}{2 \cos(\pi \alpha /2) \Gamma(n-\alpha)} 
\frac{\partial^n}{\partial x^n} 
\left(
\int^{x}_{-\infty} \frac{Z(\xi,t) d\xi}{(x-\xi)^{\alpha-n+1}}+
\int^{\infty}_x \frac{(-1)^n Z(\xi,t) d\xi}{(\xi-x)^{\alpha-n+1}}
\right) .
\ee

Multiplying Eq. (\ref{D4}) on $\exp(ikx)$, 
and integrating over $k$ from $-\infty$ to $+\infty$, we obtain
\be \label{D7}
\frac{\partial}{\partial t}Z=F(Z)+
g_0 a_{\alpha} \frac{\partial^{\alpha}}{\partial |x|^{\alpha}} Z -
2g_0 \sum^{\infty}_{n=1} \frac{\zeta(\alpha+1-2n)}{(2n)!} 
\frac{\partial^{2n}}{\partial x^{2n}} Z , \quad (\alpha \not=0,1,2...) .
\ee

The first term ($n=1$) of the sum is
$\zeta(\alpha-1) \partial^2_x Z$. 
Let us compare the coefficients  of terms with fractional and 
second derivatives in Eq. (\ref{D7}).
For $\alpha \rightarrow 2$, one can use the asymptotics
\[ \zeta(\alpha-1) \approx \frac{1}{\alpha-2}+O(1), \quad 
a_{\alpha} \approx \frac{1}{\alpha-2}+O(1)  \quad (\alpha \not=2). \]
As an example, for $\alpha=1.99$, 
\[ \zeta(\alpha-1) \approx -99.42351 , \quad a_{\alpha} \approx -100.92921  . \]
Therefore $\zeta(\alpha-1) / a_{\alpha} \sim 1$ for $2-\alpha \ll 1$.

\subsection{Infrared approximation}

In this section, we derive the main relation that permits to transfer 
the system of discrete oscillators into fractional differential equation.
This transform will be called infrared limit.
For $0<\alpha<2$, $\alpha\not=1$ and $k \rightarrow 0$, the fractional power of 
$|k|$ is a leading asymptotic term in Eq. (\ref{D4}), and
\be \label{D8}
[\tilde J_{\alpha}(0)- \tilde J_{\alpha}(k)] \approx a_{\alpha} |k|^{\alpha} ,
\quad (0<\alpha<2, \quad \alpha \not=1 ).
\ee
Eq. (\ref{D8}) can be considered as an infrared approximation of (\ref{D4}). 
Substitution of (\ref{D8}) into (\ref{C10}) gives
\be \label{D9}
\frac{\partial}{\partial t}y(k,t)=\hat F(y(k,t))-g_0 a_{\alpha} |k|^{\alpha} y(k,t),
\quad (0<\alpha<2, \quad \alpha \not=1)  .
\ee
Then 
\be \label{D10}
\frac{\partial}{\partial t}Z=F(Z)+g_0 a_{\alpha} 
\frac{\partial^{\alpha}}{\partial |x|^{\alpha}} Z ,
\quad (0<\alpha<2, \quad \alpha \not=1) , 
\ee
where $Z=Z(x,t)$ is defined by (\ref{D6}).
Eq. (\ref{D10}) can be considered as an equation for continuous 
oscillatory medium with $\alpha<2$ in infrared ($k \rightarrow 0$) 
approximation.  

As an example, for $F(z)=0$ Eq. (\ref{D10}) gives 
the fractional kinetic equation: 
\be \label{D10b}
\frac{\partial}{\partial t}Z=g_0 a_{\alpha} 
\frac{\partial^{\alpha}}{\partial |x|^{\alpha}} Z ,
\quad (0<\alpha<2, \quad \alpha \not=1 )
\ee
that describes the fractional superdiffusion \cite{SZ,Zaslavsky7,Uch}.
For $F(z)$ defined by (\ref{E2}), Eq. (\ref{D10}) is a fractional 
Ginzburg-Landau equation that has been suggested in \cite{Zaslavsky6} 
(see also \cite{Physica2005,Mil}), and will be considered in Section 3.
For $\alpha >2$ and $k \rightarrow 0$ the main term in (\ref{D2})
is proportional to $k^2$ and in (\ref{D10}), (\ref{D10b}), we have
second derivative instead of the fractional one.
The existence of critical value $\alpha=2$ was obtained in \cite{Lask}.

\section{Fractional Ginzburg-Landau (FGL) equation}

\subsection{Synchronized states for Ginzburg-Landau equation}

The one-dimensional lattice of weakly coupled nonlinear 
oscillators is described by 
\be
\frac{d}{dt} z_n(t)=(1+ia)z_n -(1+ib)|z_n|^2z_n+
(c_1+ic_2)(z_{n+1}-2z_{n}+z_{n-1}),
\ee
where we assume that all oscillators have the same parameters.
A transition to the continuous medium assumes \cite{Pik1}
that the difference $z_{n+1}-z_n$ is of order $\Delta x$, and
the interaction constants $c_1$ and $c_2$ are large.
Setting $c_1=g (\Delta x)^{-2}$, $c_2=gc(\Delta x)^{-2}$,
and $Z(x,t)\approx Z(n\Delta x,t)=z_n(t)$, we get
\be \label{A1}
\frac{\partial}{\partial t}Z=(1+ia)Z-(1+ib)|Z|^2Z+
g(1+ic)\frac{\partial^2}{\partial x^2} Z ,
\ee
which is a complex time-dependent Ginzburg-Landau  equation.
The simplest coherent structures for this equation are 
plane-wave solutions \cite{Pik1}:
\be \label{A3}
Z(x,t)=R(K)exp[ iKx-i\omega(K)t+\theta_0],
\ee
where
\be \label{A4}
R(K)=(1-gK^2)^{1/2} , \quad 
\omega(K)=(b-a)+(c-b)gK^2 ,
\ee 
and $\theta_0$ is an arbitrary constant phase.
These solutions exist for 
\be \label{A5}
gK^2<1 . \ee
Solution (\ref{A3}) can be interpreted as a synchronized state \cite{Pik1}.

\subsection{Particular solution for FGL equation}

Let us come back to the equation for
nonlinear oscillators (\ref{C1}) with $F(z)$ in Eq. (\ref{E2}) 
and long-range coupling (\ref{C2}):
\be \label{D11}
\frac{d}{dt}z_n=(1+ia)z_n-(1+ib)|z_n|^2 z_n +
g_0 \sum_{m \not=n} \frac{1}{|n-m|^{\alpha+1}} (z_n-z_m) ,
\ee
where $z_n=z_n(t)$ is the position of the $n$-th oscillator 
in the complex plane, $1<\alpha<2$.
The corresponding equation in the continuous limit and 
infrared approximation can be obtained in the 
same way as (\ref{D10}):
\be \label{A2}
\frac{\partial}{\partial t}Z=(1+ia)Z-(1+ib)|Z|^2Z+
g(1+ic)\frac{\partial^{\alpha}}{\partial |x|^{\alpha}} Z ,
\ee
where $g(1+ic)=g_0 a_{\alpha}$, and $1<\alpha<2$.
Eq. (\ref{A2}) is a fractional generalization of complex 
time-dependent Ginzburg-Landau equation (\ref{A1}) (compare to (\ref{D10})). 
Here this equation is derived in a specific approximation 
for the oscillatory medium.

We seek a particular solution of (\ref{A2}) in the form 
\be \label{A6}
Z(x,t)=A(K,t)e^{iKx} ,
\ee
which allows to use
\be \label{A7}
\frac{\partial^{\alpha}}{\partial |x|^{\alpha}} e^{iKx}=-|K|^{\alpha} e^{iKx} .
\ee
Eq. (\ref{A6}) represents a particular solution of (\ref{A2}) 
with a fixed wave number $K$.

Substitution of (\ref{A6}) into (\ref{A2}) gives
\be \label{A8}
\frac{\partial}{\partial t}A(K,t)=(1+ia)A-(1+ib)|A|^2A-
g(1+ic)|K|^{\alpha}A .
\ee
Rewriting this equation in polar coordinates,
\be \label{A9}
A(K,t)=R(K,t)e^{i\theta(K,t)},
\ee
we obtain 
\[
\frac{dR}{dt}= (1-g|K|^{\alpha})R-R^3 ,
\]
\be \label{A10}
\frac{d \theta}{dt}= (a-cg|K|^{\alpha})-b R^2.
\ee
The limit cycle here is a circle with the radius
\be \label{A11}
R=(1-g|K|^{\alpha})^{1/2} , \quad g|K|^{\alpha} <1 .
\ee
Solution of (\ref{A10}) with arbitrary initial conditions
\be \label{B1}
R(K,0)=R_0, \quad \theta(K,0)=\theta_0
\ee
is
\be \label{B2}
R(t)=R_0 (1-g|K|^{\alpha})^{1/2} 
\left( R^2_0+(1-g|K|^{\alpha}-R^2_0)e^{-2(1-g|K|^{\alpha})t} \right)^{-1/2},
\ee
\be \label{B3}
\theta(t)=
-\frac{b}{2} \ln \left[ (1-g|K|^{\alpha})^{-1} 
\left( R^2_0 +(1-g|K|^{\alpha}-R^2_0)e^{-2at} \right) \right]
-\omega_{\alpha}(K) t+\theta_0 ,
\ee
where
\be \label{B7}
\omega_{\alpha}(K)=(b-a)+(c-b)g|K|^{\alpha} ,
\quad 1-g|K|^{\alpha}>0 . 
\ee
This solution can be interpreted as a coherent structure in 
nonlinear oscillatory medium with long-range interaction.

If 
\[
R^2_0=1-g|K|^{\alpha} ,  \quad g|K|^{\alpha} <1 ,
\]
then Eqs. (\ref{B2}) and (\ref{B3}) give
\be \label{Rtheta}
R(t)=R_0, \quad \theta(t)=-\omega_{\alpha}(K) t +\theta_0 .
\ee
Solution (\ref{Rtheta}) means that
on the limit cycle (\ref{A11}) the angle variable $\theta$
rotates with a constant velocity $\omega_{\alpha}(K)$.
As the result, we have the plane-wave solution
\be \label{B6}
Z(x,t)=(1-g|K|^{\alpha})^{1/2} e^{iKx-i\omega_{\alpha}(K)t+i\theta_0} , 
\quad 1-g|K|^{\alpha}>0 ,
\ee
which can be interpreted as synchronized state
of the oscillatory medium.

For initial amplitude that deviates from (\ref{A11}), i.e.,
$R^2_0 \not= 1-g|K|^{\alpha}$, 
an additional phase shift occurs due to the term which is
proportional to $b$ in (\ref{B3}).
The oscillatory medium can be characterized by a single generalized 
phase variable.
To define it, let us rewrite (\ref{A10}) as
\be \label{A10a}
\frac{d}{dt} \ln R= (1-g|K|^{\alpha})-R^2 ,
\ee
\be \label{A10b}
\frac{d}{dt} \theta= (a-cg|K|^{\alpha})-b R^2.
\ee
Substitution of $R^2$ from (\ref{A10a}) into (\ref{A10b}) gives
\be \label{A10c}
\frac{d}{dt} \left( \theta -b \ln R \right)= (a-cg|K|^{\alpha})-b (1-g|K|^{\alpha}) .
\ee
Thus, the generalized phase \cite{Pik1} can be defined by
\be \label{A10d}
\phi(R,\theta)=\theta-b \ln R .
\ee
From (\ref{A10c}), we get
\be
\frac{d}{dt} \phi= - \omega_{\alpha}(K) .
\ee
This equation means that generalized phase $\phi(R,\theta)$ rotates uniformly 
with constant velocity.
For $g|K|^{\alpha}=(b-a)/(b-c)<1$, we have the lines of constant generalized phase.
On $(R,\theta)$ plane these lines are logarithmic spirals $\theta-b \ln R =const$.
The decrease of $\alpha$ corresponds to the increase of $K$.
For the case $b=0$ instead of spirals we have straight lines $\phi=\theta$.

\subsection{Group and phase velocity of plane-waves}

Energy propagation can be characterized by
the group velocity 
\be
v_{\alpha,g}=\frac{\partial \omega_{\alpha}(K)}{\partial K} .
\ee
From Eq. (\ref{B7}), we obtain
\be
v_{\alpha,g}=\alpha (c-b) g |K|^{\alpha-1}. 
\ee
For
\be
|K|<K_1=\left(\alpha / 2 \right)^{2-\alpha} ,
\ee
we get 
\be
|v_{\alpha,g}| > | v_{2,g}|.
\ee
The phase velocity is 
\be
v_{\alpha,ph}=\omega_{\alpha}(K) / K=(c-b)g |K|^{\alpha-1} .
\ee
For 
\be
K<K_2=2^{\alpha-2} ,
\ee
we have
\be
|v_{\alpha,ph}| > |v_{2,ph}| .
\ee
Therefore long-range interaction decreasing as $|x|^{-(\alpha+1)}$ with
$1<\alpha<2$ leads to increase the group
and phase velocities for small wave numbers ($K \rightarrow 0$).
Note that the ratio $v_{\alpha,g} /v_{\alpha,ph}$ between
the group and phase velocities of plane waves is equal to $\alpha$.

\subsection{Stability of plane-wave solution}

Solution of (\ref{B6}) can be presented as
\be \label{B11}
X=R(K,t) \cos (\theta(K,t)+Kx ), \quad 
Y=R(K,t) \sin (\theta(K,t)+Kx ),
\ee
where $X=X(K,t)=\mathrm{Re} Z(x,t)$ and $Y=Y(K,t)=\mathrm{Im} Z(x,t)$, 
and $R(K,t)$ and $\theta(K,t)$ are defined 
by (\ref{B2}) and (\ref{B3}).
For the plane-waves
\[ X_0(x,t)=(1-g|K|^{\alpha})^{1/2} 
\cos \left( Kx-\omega_{\alpha}(K)t+\theta_0 \right) , \]
\be \label{PW}
Y_0(x,t)=(1-g|K|^{\alpha})^{1/2} \sin \left( Kx-\omega_{\alpha}(K)t+\theta_0 \right) , 
\quad 1-g|K|^{\alpha}>0 .
\ee
Not all of the plane-waves are stable.
To obtain the stability condition, 
consider the variation of (\ref{A8}) near solution (\ref{PW}):
\be \label{S1}
\frac{d}{dt} \delta X(K,t)=A_{11}\delta X+A_{12}\delta Y, \quad
\frac{d}{dt} \delta Y(K,t)=A_{21}\delta X+A_{22} \delta Y ,
\ee
where $\delta X$ and $\delta Y$ are small variations of $X$ and $Y$, and
\[ A_{11}=1-g|K|^{\alpha} -2X_0(X_0-bY_0)-(X^2_0+Y^2_0), \]
\[ A_{12}=-a+gc|K|^{\alpha} -2Y_0(X_0-bY_0)+b(X^2_0+Y^2_0), \]
\[ A_{21}=a-gc|K|^{\alpha} -2X_0(Y_0+bX_0)-b(X^2_0+Y^2_0), \]
\be \label{S2}
A_{22}=1-g|K|^{\alpha} -2Y_0(Y_0+bX_0)-(X^2_0+Y^2_0) . 
\ee
The conditions of asymptotic stability for (\ref{S1}) are
\be \label{Hur}
A_{11}+A_{22}<0 , \quad A_{11}A_{22}-A_{12}A_{21}<0 .
\ee
From Eqs. (\ref{PW}) and (\ref{S2}), we get
\be \label{S7}
A_{11}+A_{22}=-2(1-g|K|^{\alpha}) , \quad
1-g|K|^{\alpha}>0 ,
\ee
and the first condition of (\ref{Hur}) is valid.
Substitution of Eqs. (\ref{PW}) and (\ref{S2}) into (\ref{Hur}) gives
\be \label{S8}
A_{11}A_{22}-A_{12}A_{21}=
\Bigl( b(1-g|K|^{\alpha})- (a-gc|K|^{\alpha} ) \Bigr)
\Bigl( 3b(1-g|K|^{\alpha})-(a-gc|K|^{\alpha} ) \Bigr) .
\ee
Then the second condition of (\ref{Hur}) has the form
\be
(V-1)(V-3)<0 ,
\ee
where
\be
V=\frac{a-gc|K|^{\alpha} }{b(1-g|K|^{\alpha})} .
\ee

As the result, we obtain
\be \label{Stab}
0<1-g|K|^{\alpha} < a/b-(c/b)g |K|^{\alpha} < 3(1-g|K|^{\alpha}) ,
\ee
i.e. the plane-wave solution (\ref{B6})
is stable if parameters $a$, $b$, $c$ and $g$ satisfy (\ref{Stab}).
Condition (\ref{Stab}) defines the region of parameters 
for plane waves where the synchronization exists.

\subsection{Forced FGL equation for isochronous case}

In this section, we consider FGL equation (\ref{A8})
forced by a constant $E$ (the so-called
forced isochronous case ($b=0$) \cite{Pik1}):
\be \label{A8e}
\frac{\partial}{\partial t}A=(1+ia)A-|A|^2A-
g(1+ic)|K|^{\alpha}A -iE, \quad (\mathrm{Im} E=0)
\ee
where $A=A(K,t)$, and we put for simplicity $b=0$, and $K$ 
is a fixed wave number. 
Our main goal will be transition to a synchronized states and 
its dependence on the order $\alpha$ of the long-range interaction.
The system of real equations is
\[
\frac{d}{dt} X=(1-g |K|^{\alpha})X -(a-gc|K|^{\alpha})Y-(X^2+Y^2)X , 
\]
\be \label{B9c}
\frac{d}{dt} Y=(1-g |K|^{\alpha})Y +(a-gc|K|^{\alpha})X-(X^2+Y^2)Y-E ,
\ee 
where $X=X(K,t)$ is real and $Y=Y(K,t)$ imaginary parts of $A(K,t)$.

Numerical solution of Eq. (\ref{B9c}) was performed with parameters
$a=1$, $g=1$, $c=70$, $E=0.9$, $K=0.1$, 
for $\alpha$ within interval $\alpha \in(1;2)$.
The results are presented on Fig. 2, and Fig. 3.
For $\alpha_0< \alpha<2$, where $\alpha_0\approx 1.51...$,  
the only stable solution is a stable fixed point.
This is region of perfect synchronization (phase locking), 
where the synchronous oscillations have a constant amplitude
and a constant phase shift with respect to the external force.
For $\alpha<\alpha_0$ the global attractor for (\ref{B9c})
is a limit cycle.
Here the motion of the forced system is quasiperiodic.
For $\alpha=2$ there is a stable node.
When $\alpha$ decreases, the stable mode transfers into
a stable focus.
At the transition point it loses stability, and 
a stable limit cycle appears.
As the result, we have that the decrease of 
order $\alpha$ from 2 to 1 
leads to the loss of synchronization (see Figures 2-3).

In Fig. 2, ($\alpha=2.00$, $\alpha=1.70$, and $\alpha=1.60$, $\alpha=1.56$),
we see that in the synchronization region
all trajectories are attracted to a stable node.

In Fig. 3, ($\alpha=1.54$, $\alpha=1.52$, and $\alpha=1.50$, $\alpha=1.40$),
a stable limit cycle appears via the Hopf bifurcation.
For $\alpha=1.54$, and $\alpha=1.52$, near the boundary of 
synchronization the fixed point is a focus.
For $\alpha=1.4$, the amplitude of the limit cycle grows, 
and synchronization breaks down.


\begin{figure}
\centering
\rotatebox{270}{\includegraphics[width=7 cm,height=7 cm]{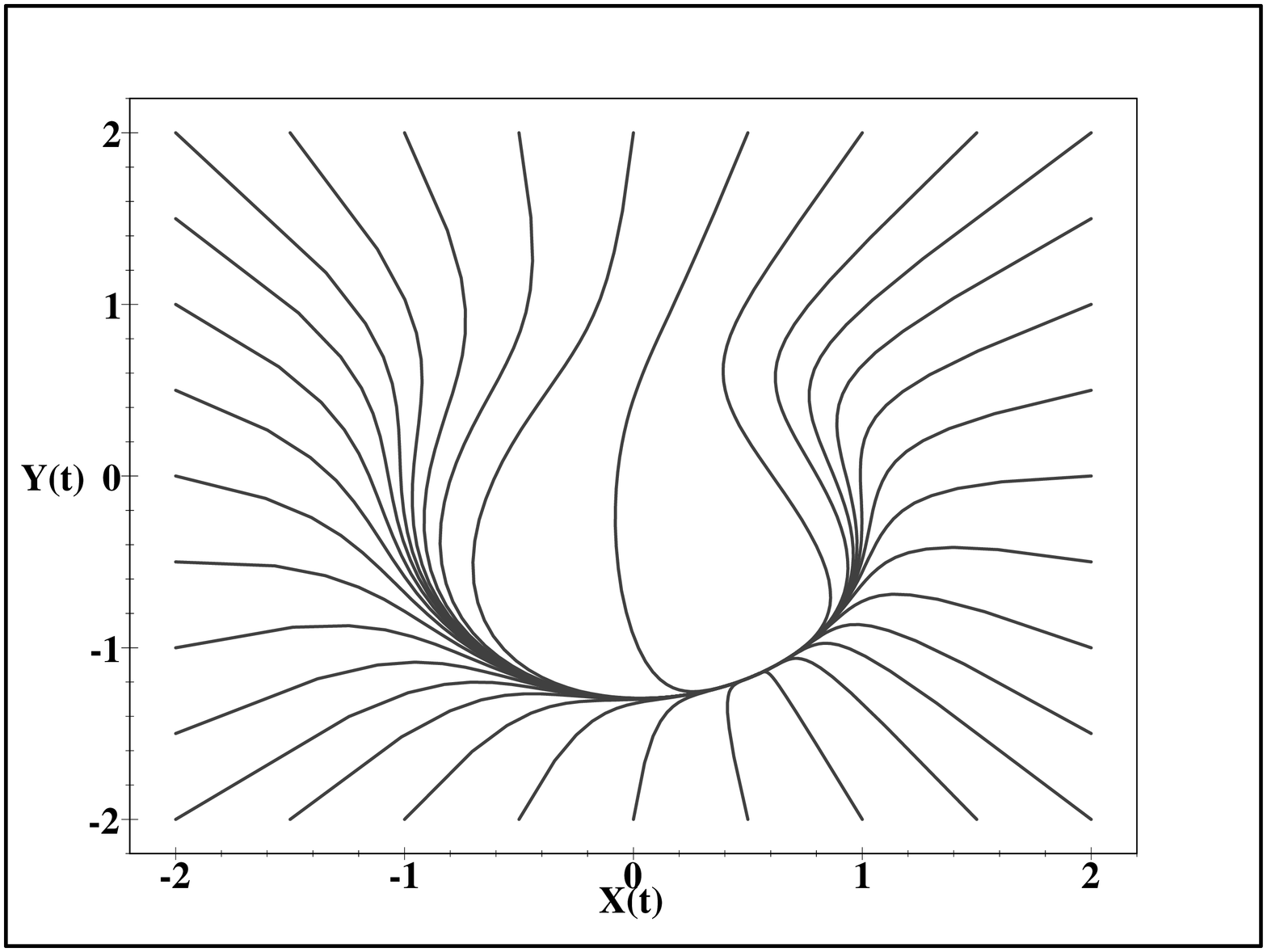}}
\rotatebox{270}{\includegraphics[width=7 cm,height=7 cm]{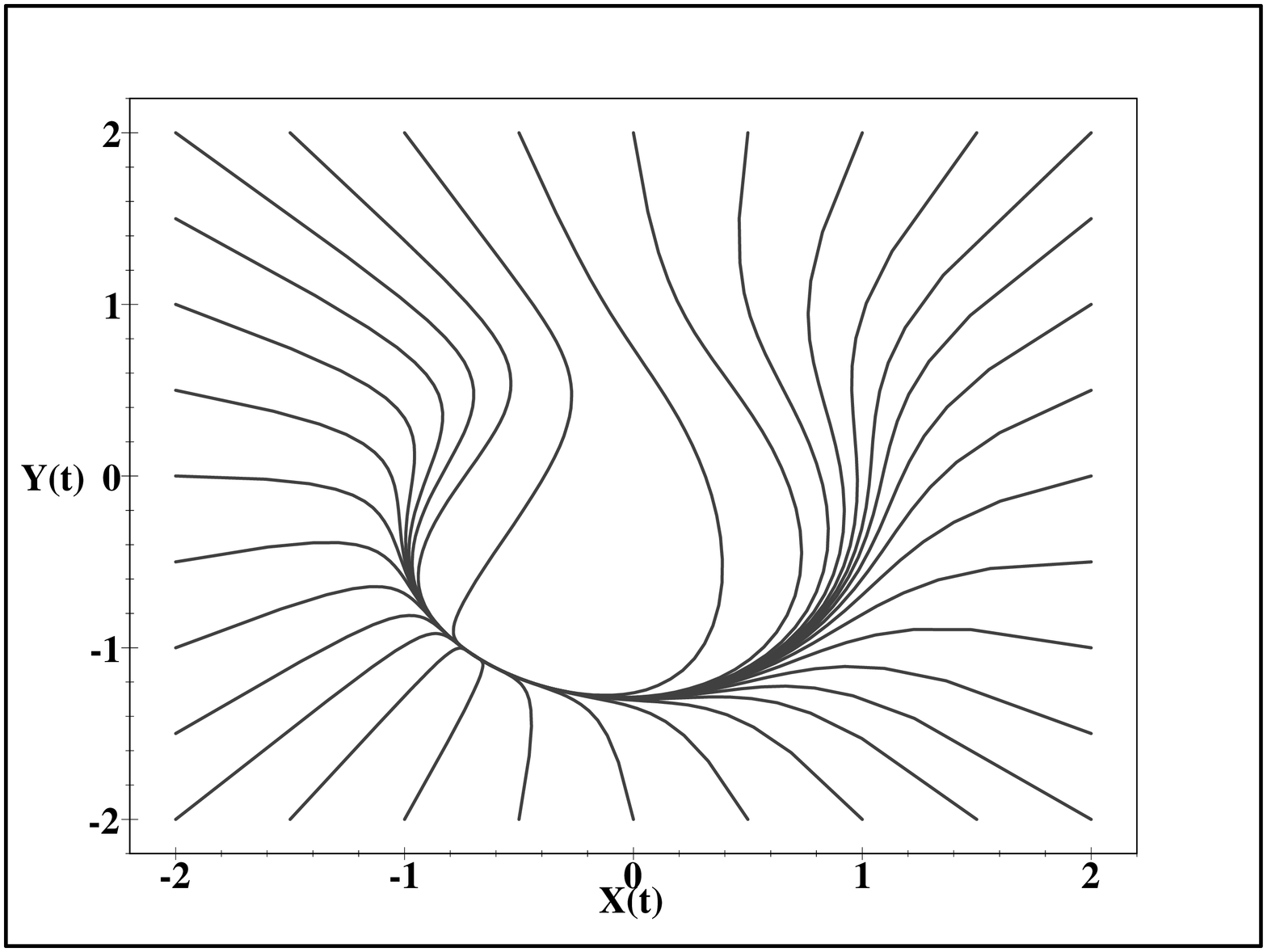}}
\rotatebox{270}{\includegraphics[width=7 cm,height=7 cm]{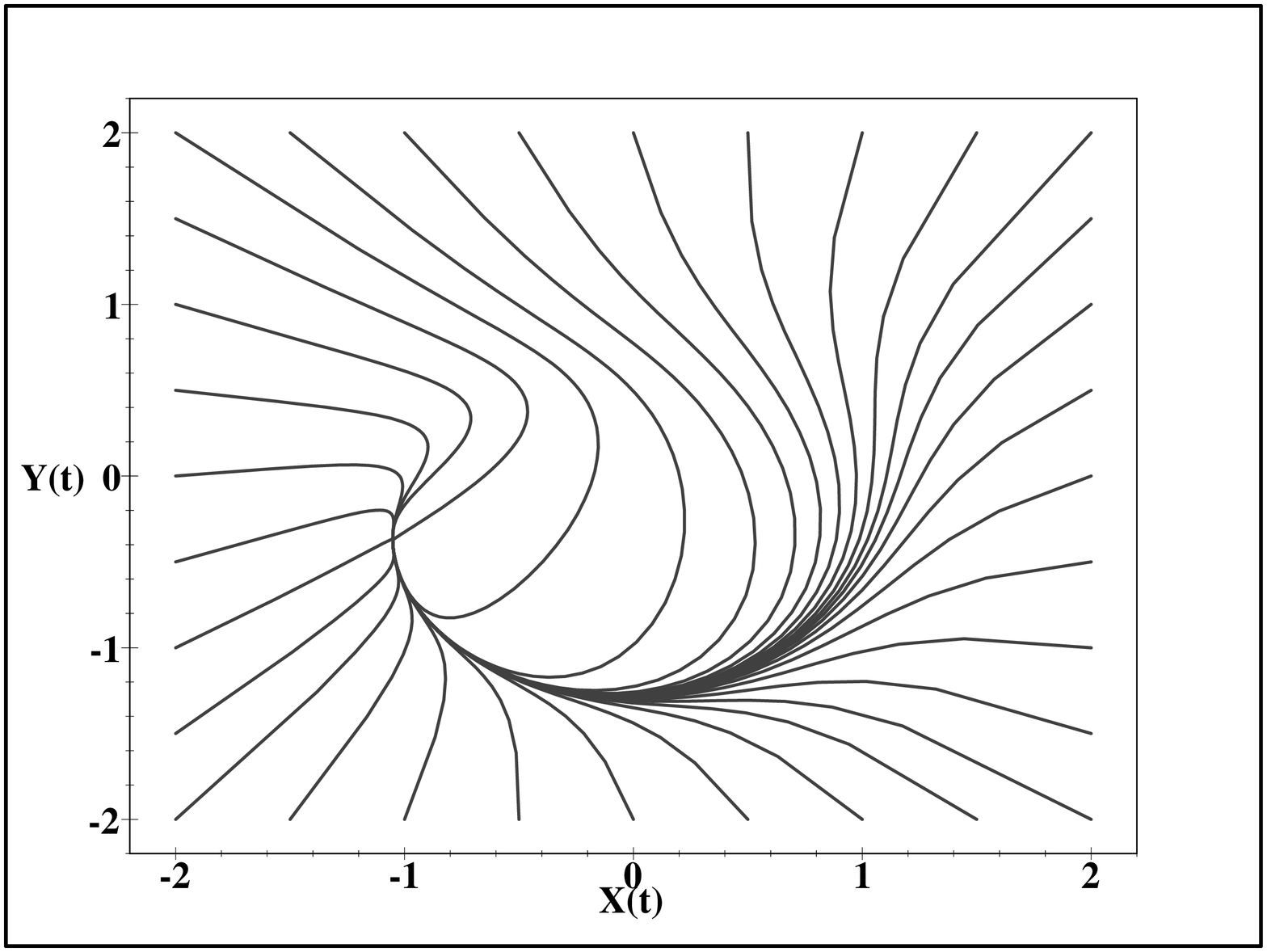}}
\rotatebox{270}{\includegraphics[width=7 cm,height=7 cm]{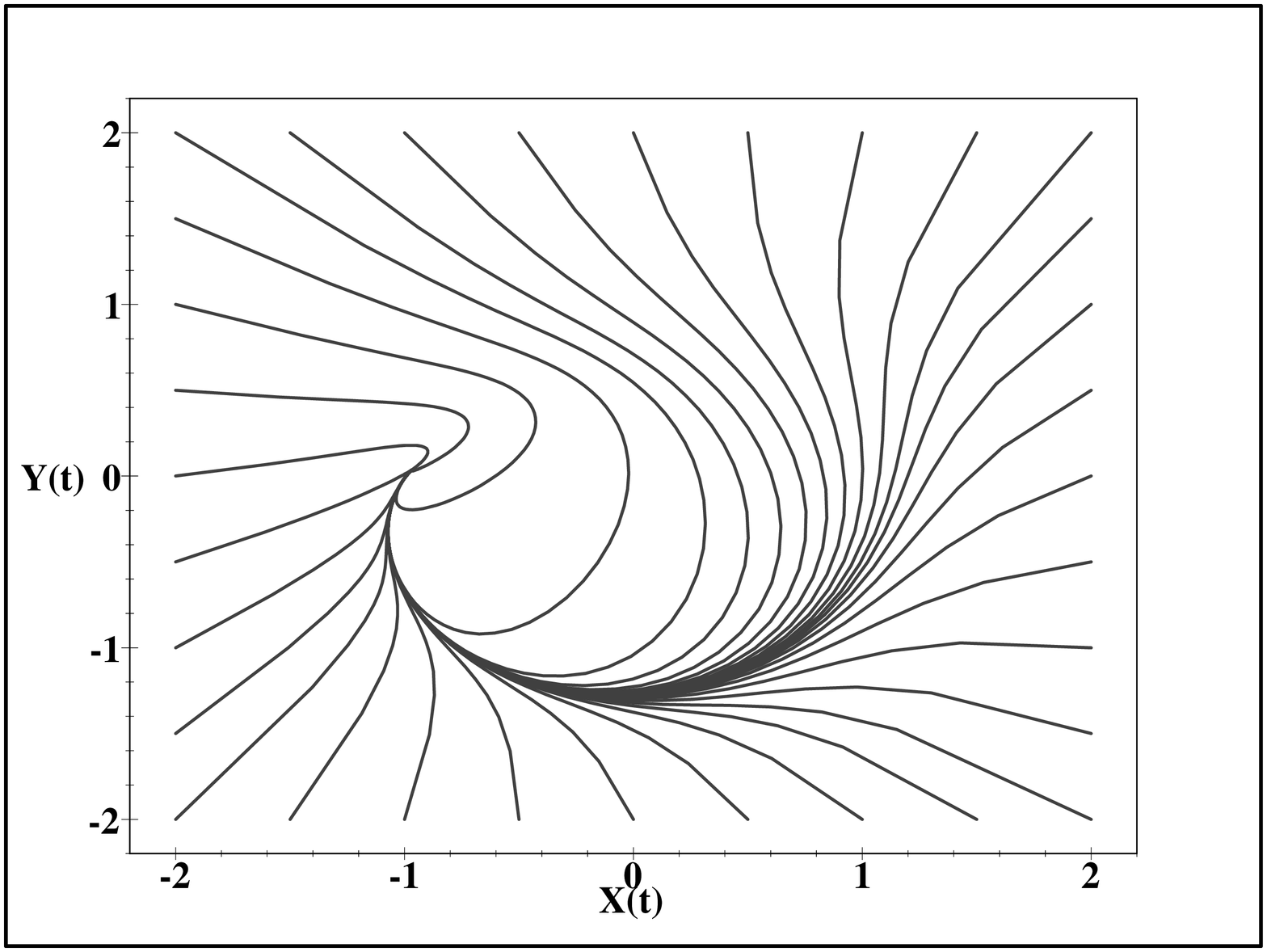}}
\caption{\label{fig4a} 
Approaching to the bifurcation point $\alpha=\alpha_0=1.51...$ of
solution of forced FGL equation for isochronous case
with fixed wave number $K=0.1$ is represented by real 
$X(K,t)$ and imaginary $Y(K,t)$ parts of $A(K,t)$.
The plots for orders 
$\alpha=2.00$, $\alpha=1.70$, $\alpha=1.60$, $\alpha=1.56$.
}
\end{figure}

\begin{figure}
\centering
\rotatebox{270}{\includegraphics[width=7 cm,height=7 cm]{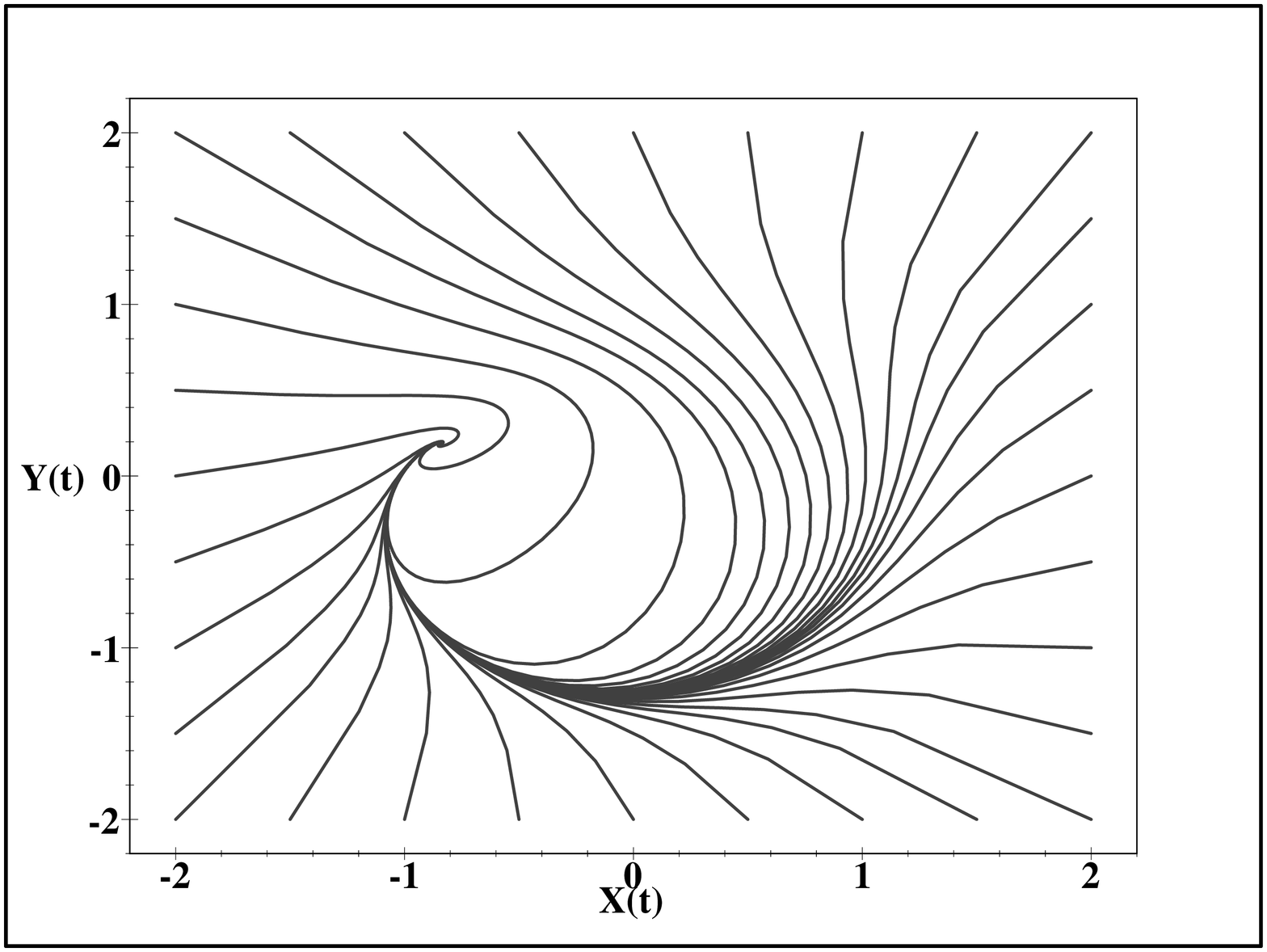}}
\rotatebox{270}{\includegraphics[width=7 cm,height=7 cm]{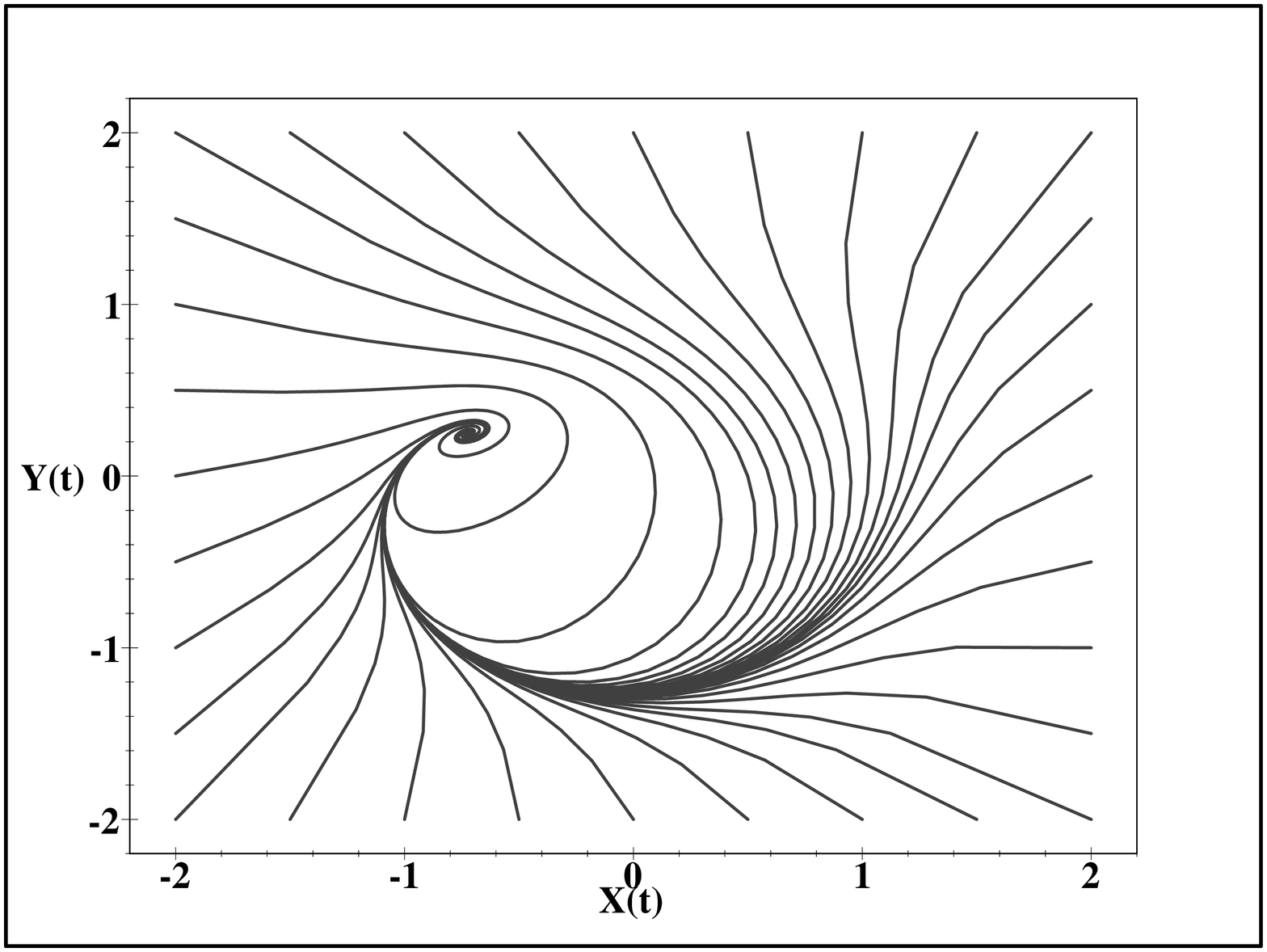}}
\rotatebox{270}{\includegraphics[width=7 cm,height=7 cm]{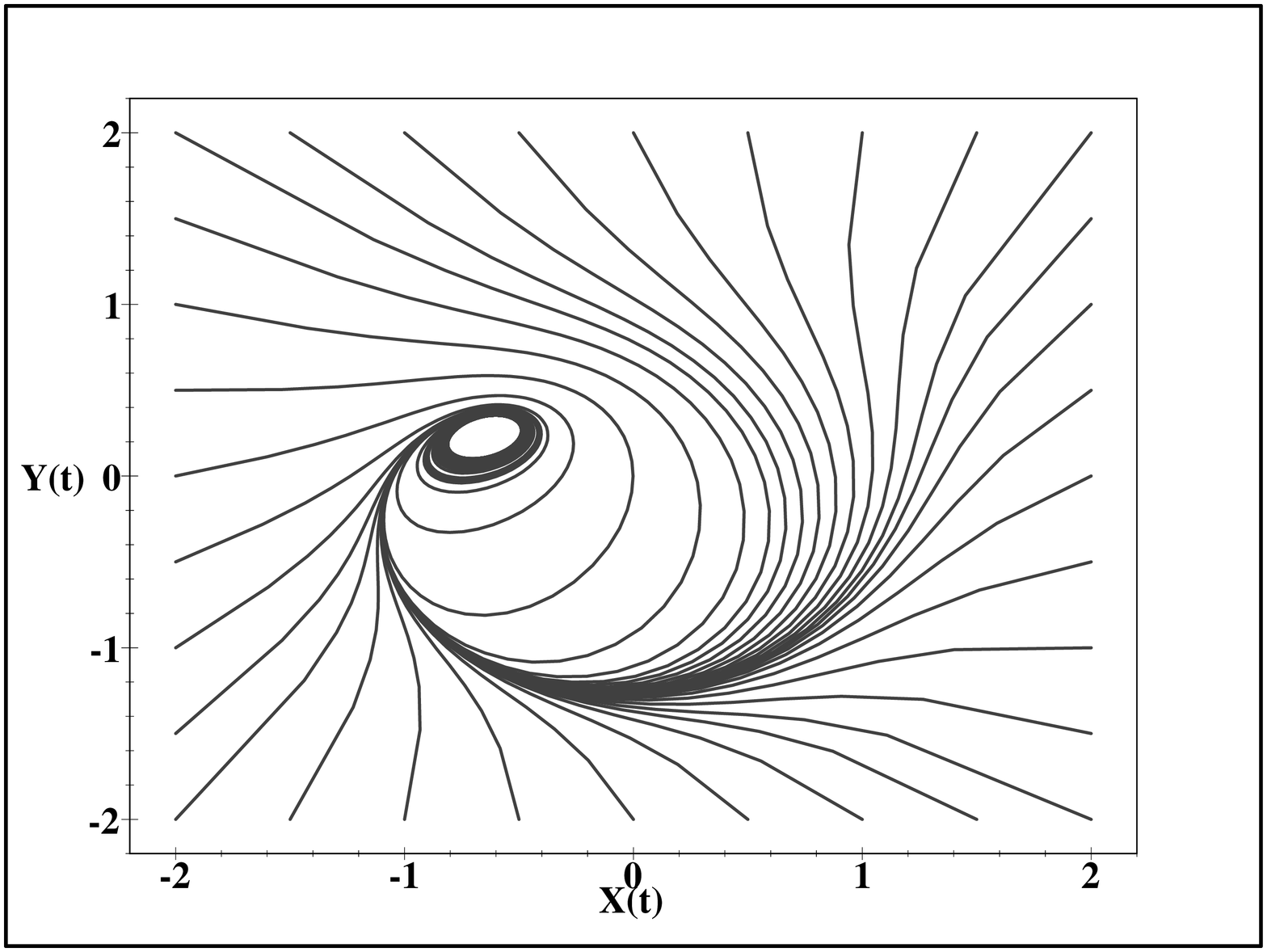}}
\rotatebox{270}{\includegraphics[width=7 cm,height=7 cm]{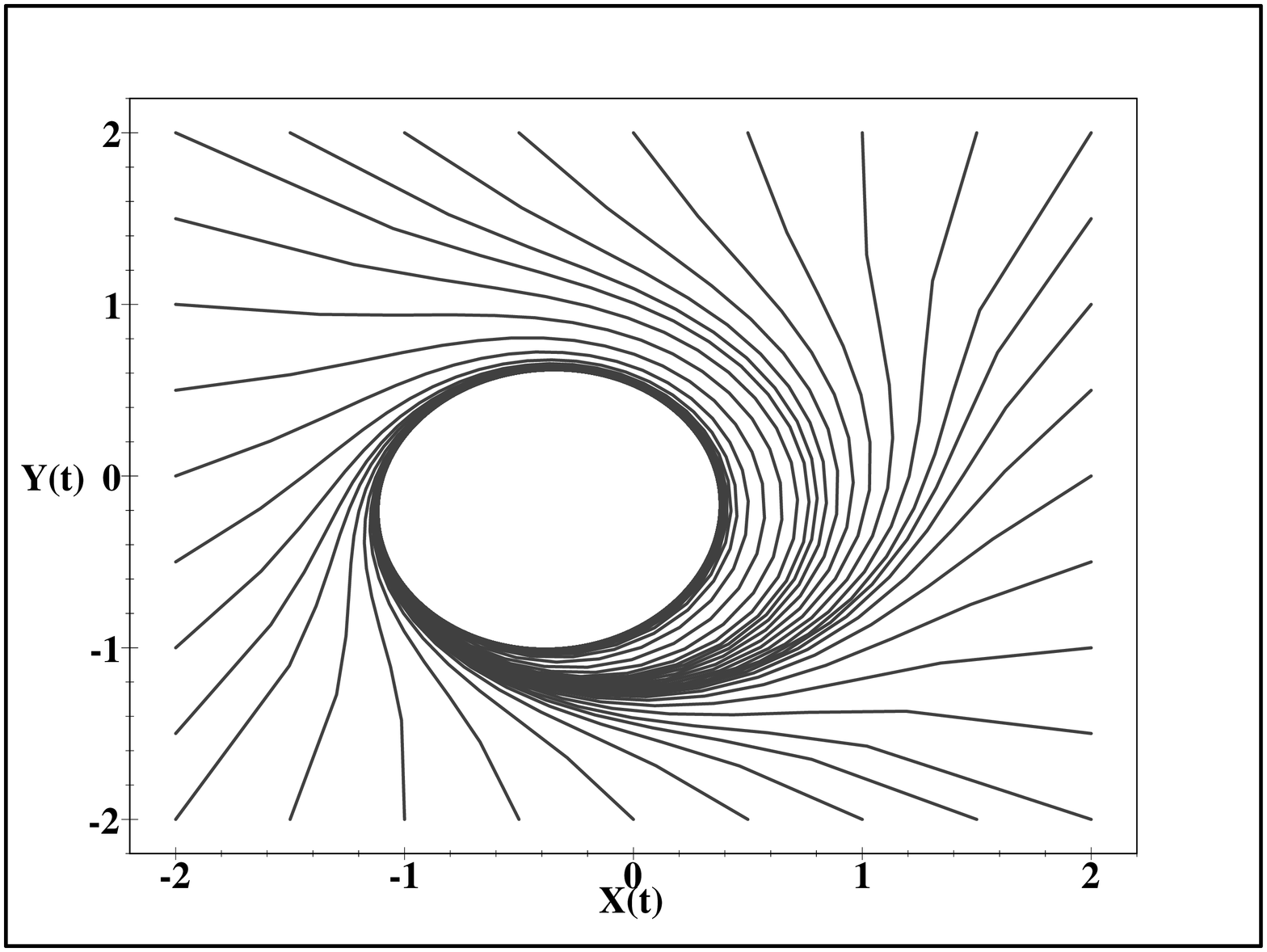}}
\caption{\label{fig4b} 
Transformation to the limit cycle of
solution of forced FGL equation for isochronous case
with fixed wave number $K=0.1$ is represented by real
$X(K,t)$ and imaginary $Y(K,t)$ parts of $A(K,t)$.
The plots for orders 
$\alpha=1.54$, $\alpha=1.52$, $\alpha=1.50$, $\alpha=1.40$.
}
\end{figure}

\subsection{Phase and amplitude for forced FGL equation}

The oscillator medium can be characterized by a single generalized phase 
variable (\ref{A10d}). We can rewrite (\ref{A10d}) as
\be \label{phiXY}
\phi(X,Y)=\arctan (Y/X) -\frac{b}{2} \ln (X^2+Y^2) , 
\ee
where $X$ and $Y$ are defined by (\ref{B11}).
For $E=0$, the phase rotates uniformly
\be
\frac{d}{dt} \phi= - \omega_{\alpha}(K)= a-gc|K|^{\alpha}, 
\ee
where $\omega_{\alpha}(K)$ is gived by (\ref{B7}) with $b=0$, and
can be considered as a frequency of natural oscillations.
For $E \not=0$, Eqs. (\ref{B9c}) and (\ref{phiXY}) give
\be
\frac{d}{dt} \phi= - \omega_{\alpha}(K) -E \cos \phi .
\ee
This equation has an integral of motion.
The integral is
\be
I_1=2(\omega^2-E^2)^{-1/2} \arctan \Bigl( 
(\omega-E) (\omega^2-E^2)^{-1/2} \tan (\phi(t)/2) \Bigr)+t, \quad
\omega^2>E^2 ,
\ee
\be
I_2=2(E^2-\omega^2)^{-1/2} \mathrm{arctanh} \Bigl( 
(E-\omega) (E^2-\omega^2)^{-1/2} \tan (\phi(t)/2) \Bigr)+t, \quad
\omega^2<E^2 .
\ee
These expressions help to obtain the solution in form (\ref{A9}) for
forced case (\ref{A8e}) keeping the same notations as in (\ref{A9}).
For polar coordinates we get 
\[
\frac{dR}{dt}= (1-g|K|^{\alpha})R-R^3-E \sin \theta ,
\]
\be \label{A10e}
\frac{d \theta}{dt}= (a-cg|K|^{\alpha})-\frac{E \cos \theta}{R}.
\ee
Numerical solution of (\ref{A10e}) was performed with 
the same parameters as for Eq. (\ref{B9c}), i.e.,
$a=1$, $g=1$, $c=70$, $E=0.9$, $K=0.1$, and $\alpha$ 
within interval $\alpha \in(1,2)$. 
The results are presented in Fig. 4 and Fig. 5.

The time evolition of phase $\theta(K,t)$ is given in Fig. 4
for $\alpha=2.00$, $\alpha=1.50$, $\alpha=1.47$, 
$\alpha=1.44$, $\alpha=1.40$, $\alpha=1.30$, $\alpha=1.20$, 
$\alpha=1.10$. The decrease of $\alpha$ from $2$ to $1$ leads
to the oscillations of the phase $\theta (K,t)$ after the Hopf 
bifurcation at $\alpha_0=1.51...$, then 
the amplitude of phase oscillation decreases and
the velocity of phase rotations increases.

The amplitude $R(K,t)$ is shown on Fig. 5 for
$ \alpha= 1.6$,  $\alpha=1.55$, $\alpha=1.55$, $\alpha=1.51$,
$\alpha=1.50$, $\alpha=1.45$, $\alpha=1.2$.  
The appearance of oscillations in the plots
means the loss of synchronization.


\begin{figure}
\centering
\rotatebox{270}{\includegraphics[width=15 cm,height=15 cm]{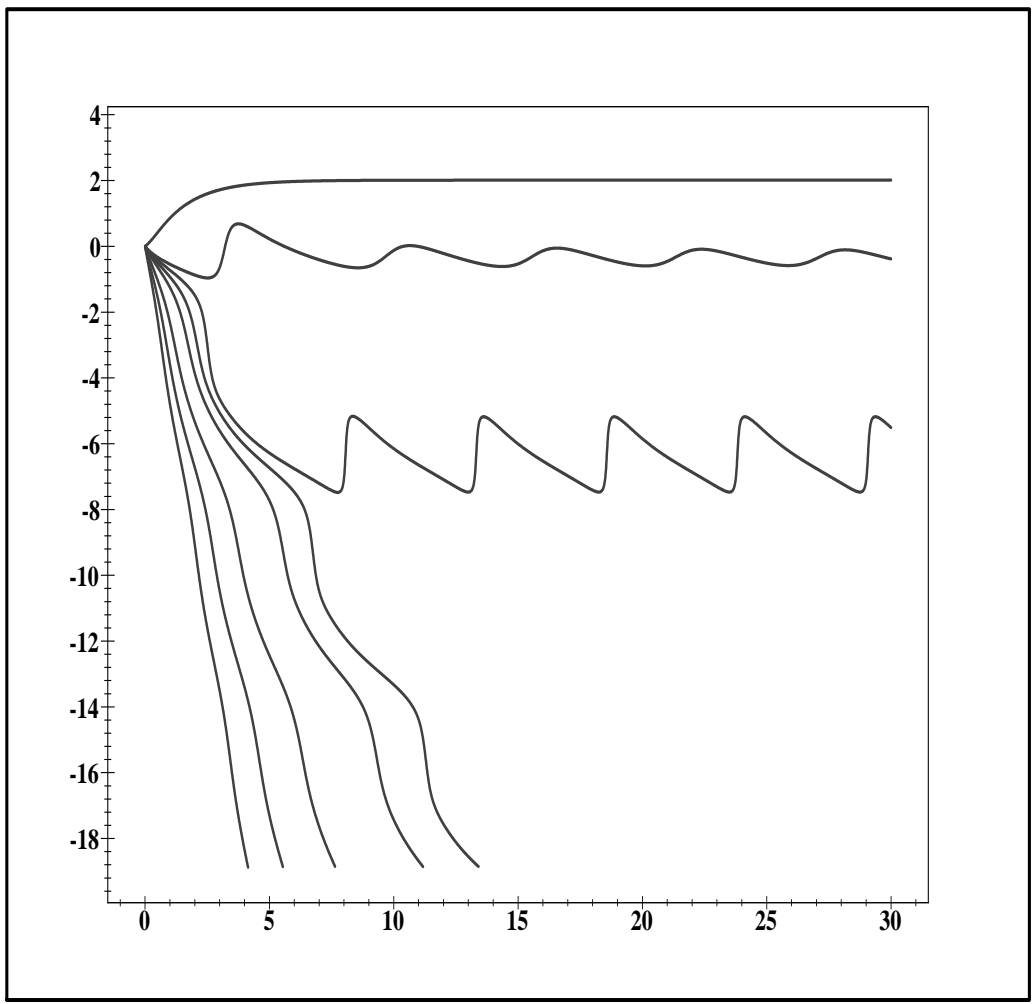}}
\caption{\label{fig2} Phase $\theta(K,t)$ for $K=0.1$ and
$\alpha= 2.00$, $\alpha=1.50$, $\alpha=1.47$, $\alpha=1.44$, 
$\alpha=1.40$, $\alpha=1.30$, $\alpha=1.20$, $\alpha=1.10$.
The decrease of order $\alpha$ corresponds to the clockwise rotation of curves. 
For upper curve $\alpha=2$. For the most vertical curve $\alpha=1.1$. }
\end{figure}

\begin{figure}
\centering
\rotatebox{270}{\includegraphics[width=6.7 cm,height=6.7 cm]{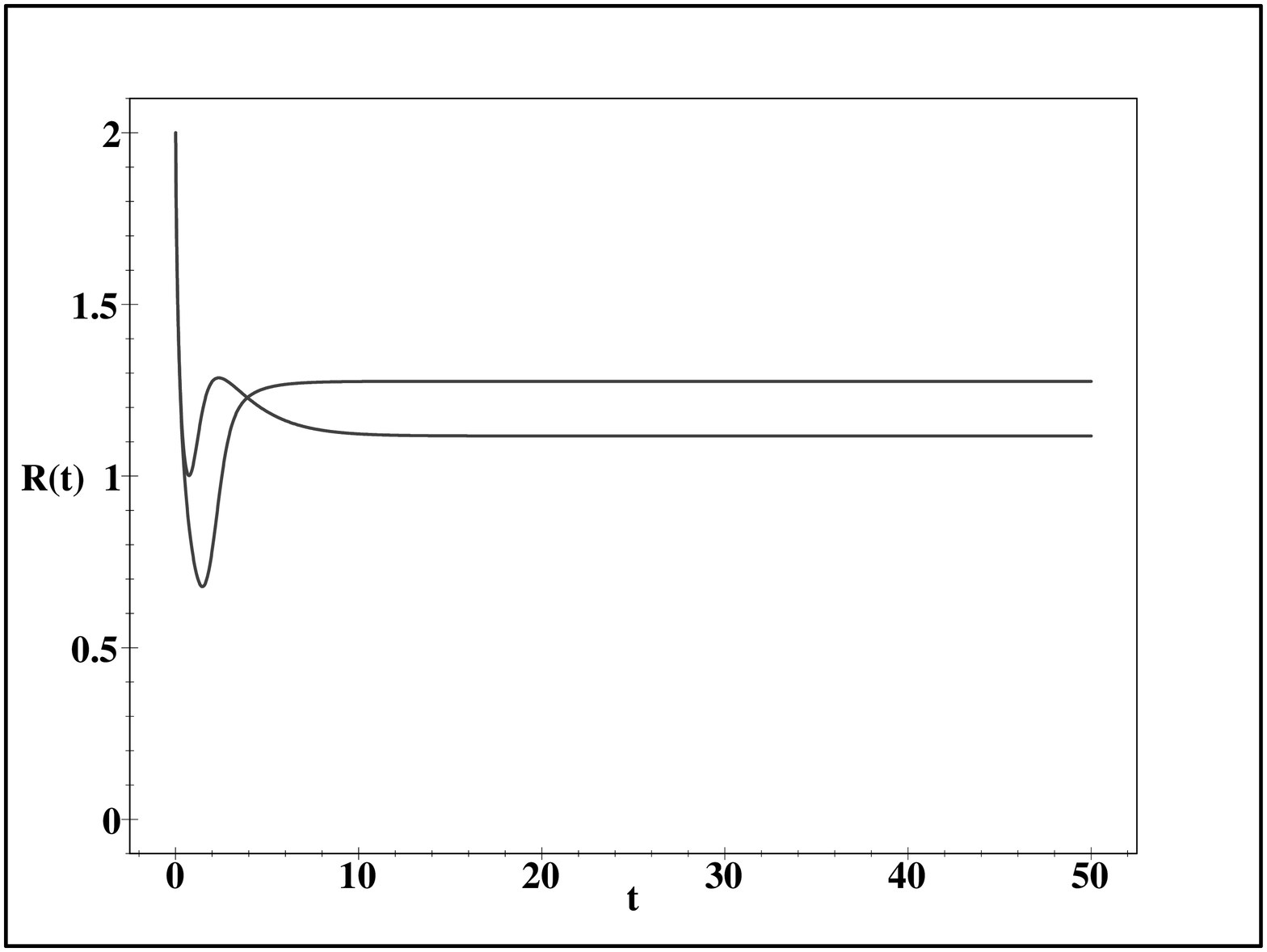}}
\rotatebox{270}{\includegraphics[width=6.7 cm,height=6.7 cm]{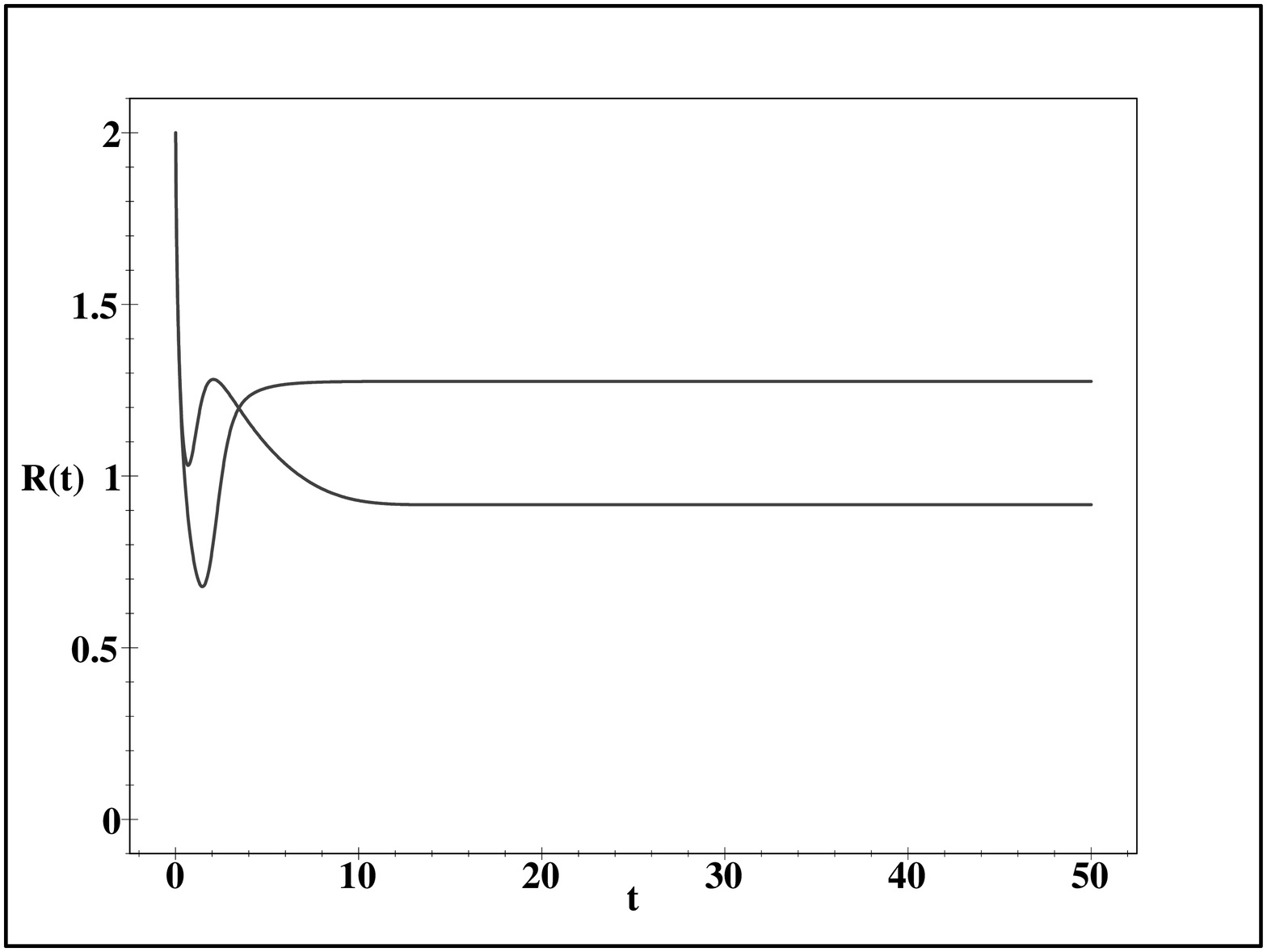}}
\rotatebox{270}{\includegraphics[width=6.7 cm,height=6.7 cm]{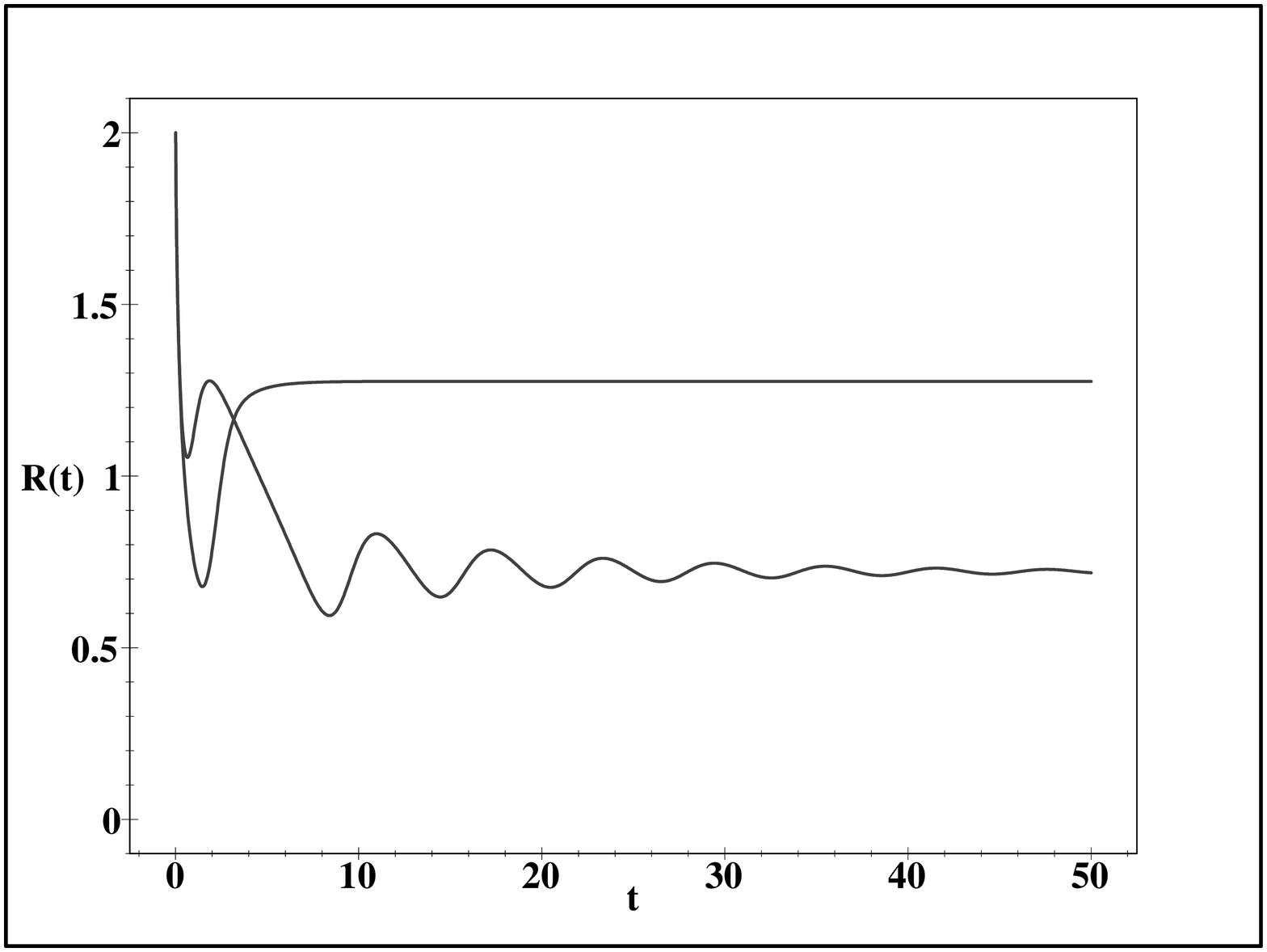}}
\rotatebox{270}{\includegraphics[width=6.7 cm,height=6.7 cm]{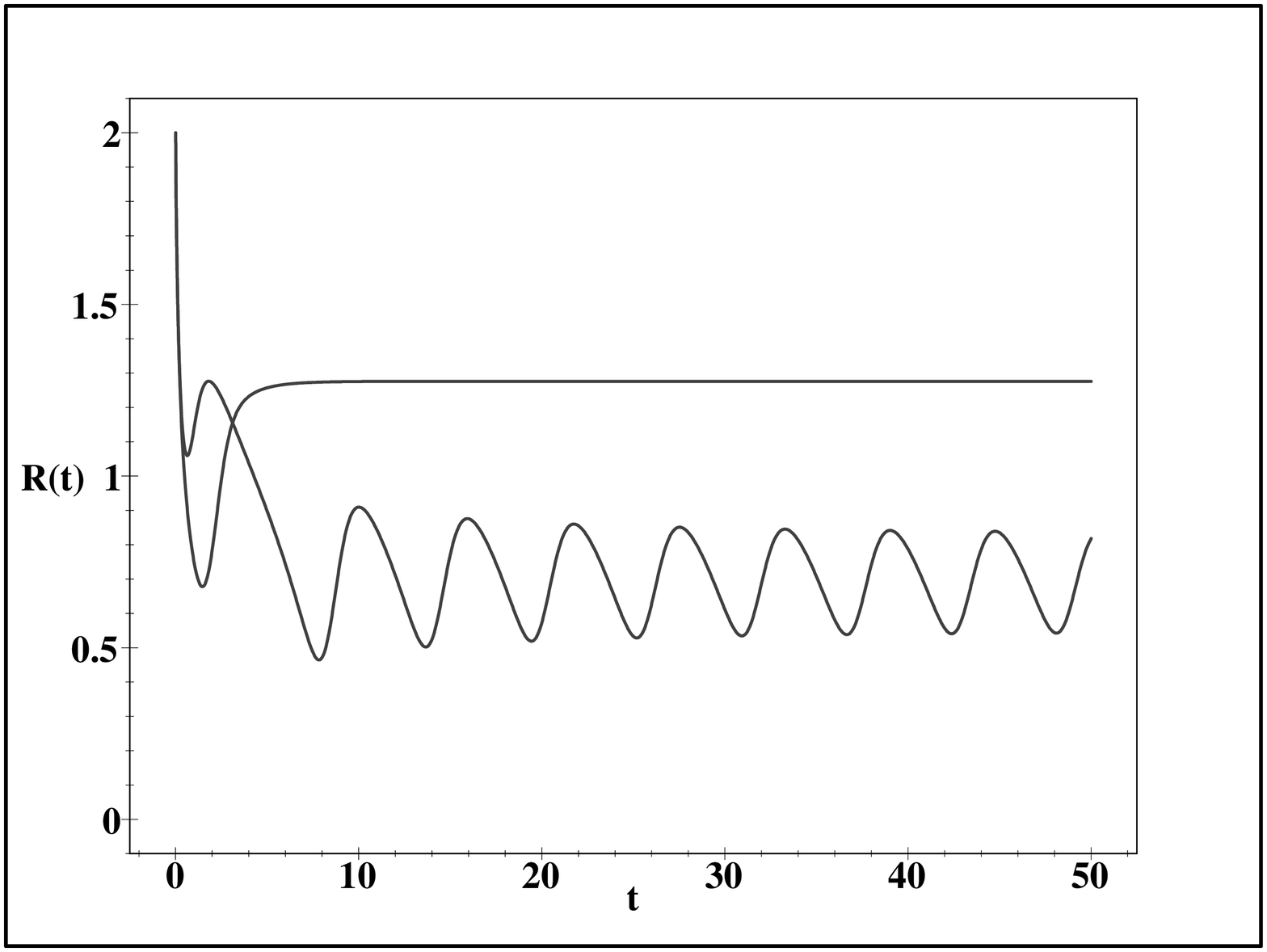}}
\rotatebox{270}{\includegraphics[width=6.7 cm,height=6.7 cm]{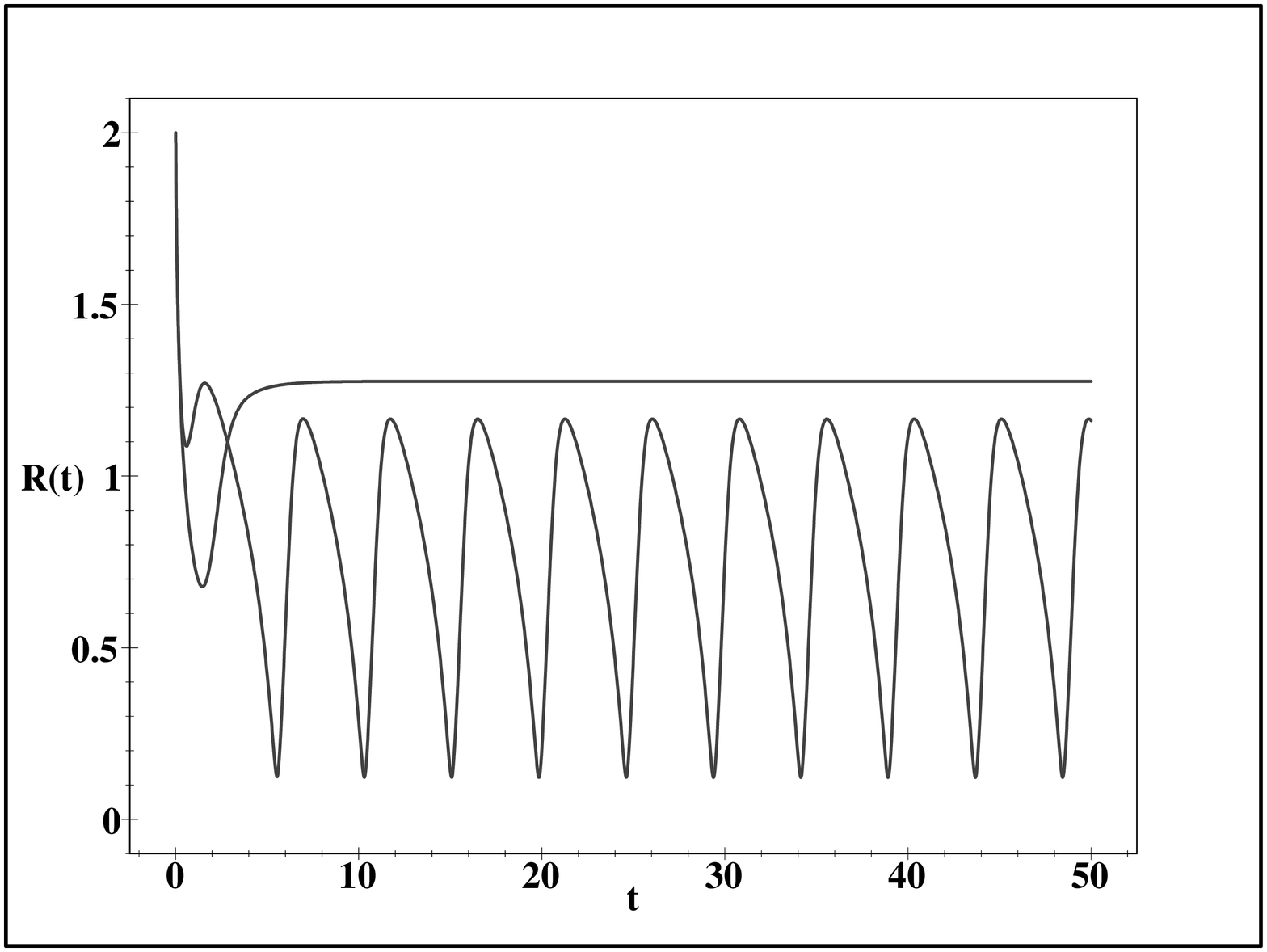}}
\rotatebox{270}{\includegraphics[width=6.7 cm,height=6.7 cm]{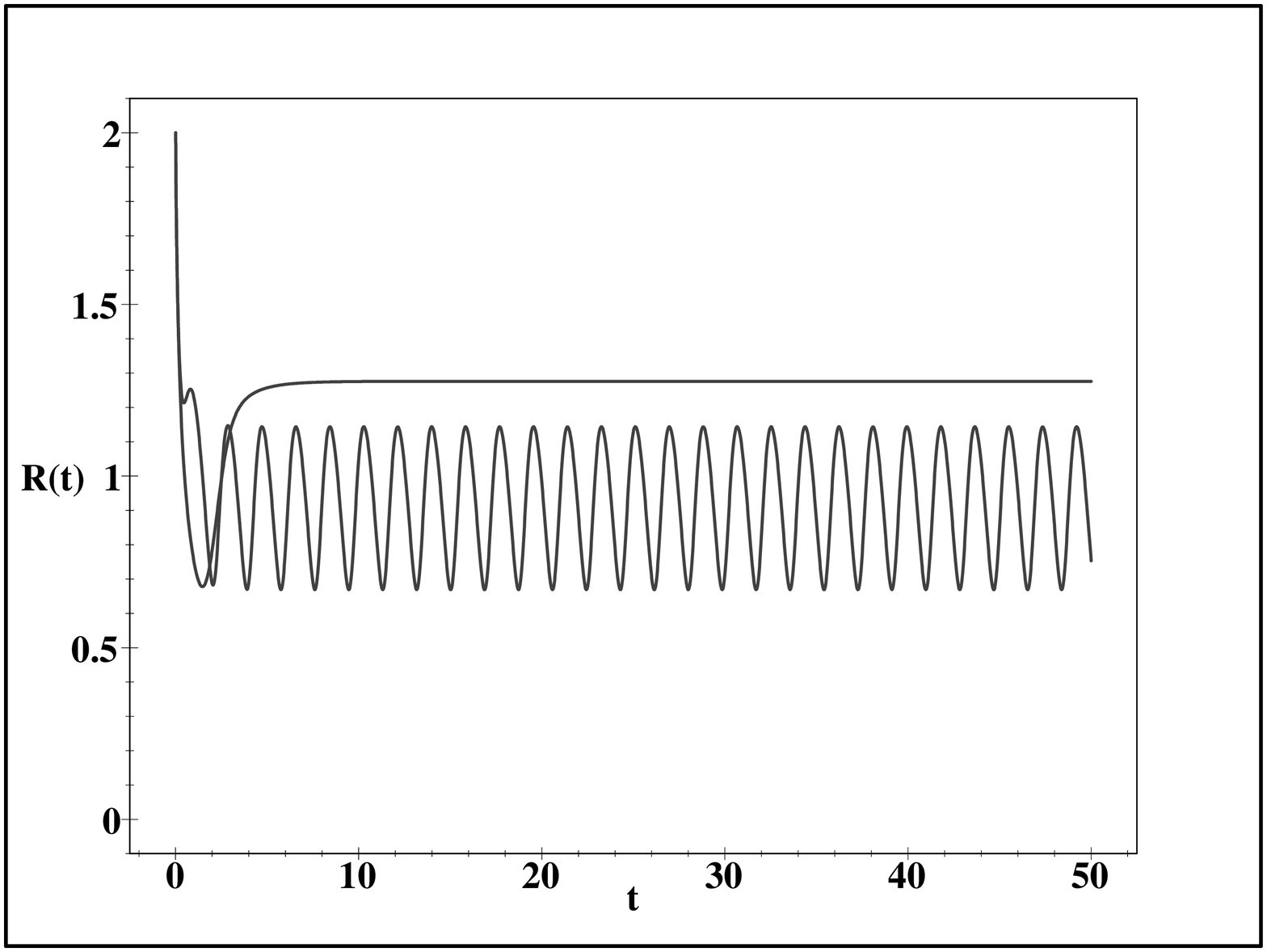}}
\caption{\label{fig3} 
Amplitude $R(K,t)$. The upper curve corresponds to $\alpha=2$ for all plots. 
The lower curves correspond to $\alpha=1.6$, $\alpha=1.55$, $\alpha=1.51$,
$\alpha=1.50$, $\alpha=1.45$, $\alpha=1.2$.  
The appearance of oscillations on the plots
means the loss of synchronization.}
\end{figure}

\section{Space-structures from FGL equation}

In previous sections, we considered mainly time-evolution and 
"time-structures" as solutions for the FGL equation. 
Particularly, synchronization process was an example of the
solution that converged to a time-coherent structure.
Here we focuse on the space structures for the solution
of FGL equation (\ref{A2}) with $b=c=0$ and the 
constants $a_1$ and $a_2$ ahead of linear term:
\be \label{H1}
\frac{\partial}{\partial t}Z=(a_1+ia_2)Z-|Z|^2Z+
g\frac{\partial^{\alpha}}{\partial |x|^{\alpha}} Z .
\ee

Let us seek a particular solution of (\ref{H1}) in the form 
\be \label{H2}
Z(x,t)=R(x,t)e^{i\theta(t)} , \quad R^*(x,t)=R(x,t),  \quad \theta^*(t)=\theta(t) .
\ee
Substitution of (\ref{H2}) into (\ref{H1}) gives
\be \label{H3}
\frac{\partial}{\partial t}R +i R \frac{\partial}{\partial t}\theta(t)
=(a_1+ia_2)R-R^3+ g \frac{\partial^{\alpha}}{\partial |x|^{\alpha}} R ,
\ee
or
\be \label{H4}
\frac{\partial}{\partial t}R=
a_1R-R^3- g\frac{\partial^{\alpha}}{\partial |x|^{\alpha}} R , \quad
\frac{\partial}{\partial t}\theta(t)=a_2 .
\ee
Using $\theta(t)=a_2t+\theta(0)$, we arrive to the existence of a 
limit cycle  with $R_0=a^{1/2}_1 $.

Let us find a particular solution of (\ref{H4}) in the vicinity 
of the limit cycle, i.e.
\be \label{H9}
R(x,t)=R_0+\varepsilon R_1+\varepsilon^2 R_2+... , \quad
(\epsilon \ll 1) .
\ee
Zero approximation $R_0=a^{1/2}_1$ satisfies (\ref{H4}) since
\be \label{H11}
\frac{\partial^{\alpha}}{\partial |x|^{\alpha}} 1=0 ,
\ee
and for $R_1=R_1(x,t)$, we have
\be \label{HH1}
\frac{\partial}{\partial t}R_1=
-2a_1R_1+ g\frac{\partial^{\alpha}}{\partial |x|^{\alpha}}  R_1 .
\ee

Consider the Cauchy problem for (\ref{HH1}) with initial condition 
\be \label{HH2}
R_1 (x,0)=\varphi(x).
\ee
By solution of the Cauchy problem we mean a function 
$R_1(x,t)$ which satisfies (\ref{HH1}) and condition (\ref{HH2}).
The Green function (or fundamental solution)
of the Cauchy problem is the (generalized) function
$G(x,t)$ such that
\be \label{HH3}
R_1(x,t)=\int^{+\infty}_{-\infty} G(x^{\prime},t) \varphi(x- x^{\prime}) dx^{\prime} .
\ee
The Green function is the solution of (\ref{HH1}) with $\varphi(x)=\delta(x)$.

Let us apply Laplace transform for $t$ and Fourier transform for $x$:
\be \label{HH4}
\tilde G(k,s)=\int^{\infty}_0 dt 
\int^{+\infty}_{-\infty} dx \ e^{-st+ikx} G(x,t) .
\ee
By the definition of Riesz derivative,
\be \label{HH5}
\frac{\partial^{\alpha}}{\partial |x|^{\alpha}} G(x,t)
\longleftrightarrow -|k|^{\alpha} \tilde G(k,s),
\ee
and for the Laplace transform with respect to time 
\be \label{HH6}
\frac{\partial}{\partial t} G(x,t)
\longleftrightarrow s \tilde G(k,s)-1.
\ee
Applying (\ref{HH4})-(\ref{HH6}) to (\ref{HH1}), we obtain
\be  \label{HH7}
s \tilde G(k,s)-1=-2a_1 \tilde G(k,s)-g |k|^{\alpha} \tilde G(k,s) ,
\ee
or 
\be  \label{HH8}
\tilde G(k,s)=\frac{1}{s+2a_1+ g |k|^{\alpha} }.
\ee
Let us first invert the Laplace transform in (\ref{HH8}):
\be \label{HH9}
\frac{1}{s+a} \longleftrightarrow e^{-at} ,
\ee
Then, the Fourier transform of the Green function 
\be
\hat G(k,t) = \int^{+\infty}_{-\infty} dx \ e^{ikx} G(x,t)
\ee
has the form
\be \label{HH10}
\hat G(k,t)=e^{ -(2a_1+g|k|^{\alpha}) t }=e^{-2a_1t} e^{ -g|k|^{\alpha} t} .
\ee
As the result, we get
\be \label{HH11}
G(x,t)=(gt)^{-1/\alpha} e^{-2a_1t}  L_{\alpha} ( x (gt)^{-1/\alpha} ) .
\ee 
where 
\be
L_{\alpha}(x)=\frac{1}{2\pi} \int^{+\infty}_{-\infty} dk \  e^{-ikx} e^{-a|k|^{\alpha}} 
\ee
is a Levy stable p.d.f. \cite{Feller}.
The p.d.f. $L_{\alpha}(x)$ for $\alpha=2.0$, 
$\alpha=1.6$, and $\alpha=1.0$ are shown on Fig. 6.

As an example, for $\alpha=1$ we have
the Cauchy distribution with respect to coordinate
\be  \label{HHH1}
e^{-|k|} \longleftrightarrow L_1 (x)=\frac{1}{\pi} \frac{1}{x^2+1} ,
\ee
and
\be \label{HHH2}
G(x,t)=\frac{1}{\pi} 
\frac{ (gt)^{-1} e^{-2a_1t}}{x^2 (gt)^{-2} +1}  .
\ee 
For $\alpha=2$, we get the Gauss distribution: 
\be  \label{HHH3}
e^{-k^2} \longleftrightarrow L_2 (x)= \frac{1}{2\sqrt{\pi}} e^{-x^2/4} ,
\ee
and
\be \label{HHH4}
G(x,t)=(gt)^{-1/2} e^{-2a_1t} \frac{1}{2\sqrt{\pi}} 
e^{-x^2 /(4gt) } .
\ee 

For $1< \alpha \le 2$ the function $L_{\alpha}(x)$ can be presented  
as the convergent expansion
\be
L_{\alpha}(x)=-\frac{1}{\pi x} \sum^{\infty}_{n=1} 
(-x)^n \frac{\Gamma(1+n/\alpha)}{n!} \sin (n \pi/2) .
\ee
The asymptotic ($x \rightarrow \infty$, $1<\alpha<2$) is given by
\be
L_{\alpha}(x) \sim -\frac{1}{\pi x} \sum^{\infty}_{n=1} 
(-1)^n x^{-n \alpha} \frac{\Gamma(1+n \alpha)}{n!} \sin (n \pi/2)  , 
\quad x \rightarrow \infty ,
\ee
with the leading term: 
\be
L_{\alpha}(x) \sim \pi^{-1} \Gamma(1+ \alpha) x^{-\alpha-1} , 
\quad x \rightarrow \infty .
\ee

As the result, the solution of (\ref{H1}) is
\be \label{SolZxt}
Z(x,t)=e^{i(a_2t+\theta(0))} \left( a^{1/2}_1+ \varepsilon 
(gt)^{-1/\alpha} e^{-2a_1t} \int^{+\infty}_{-\infty} 
L_{\alpha} ( x^{\prime} (gt)^{-1/\alpha} ) \
\varphi(x- x^{\prime}) dx^{\prime} +O(\varepsilon^2) \right) .
\ee
This solution can be considered as a space-time synchronization 
in the oscillatory medium with long-range interaction 
decreasing as $|x|^{-(\alpha+1)}$.

For $\varphi(x)=\delta(x-x_0)$, solution (\ref{SolZxt}) has the form
\be
Z(x,t)=e^{i(a_2t+\theta(0))} \Bigl( 
a^{1/2}_1 + \varepsilon (gt)^{-1/\alpha} e^{-2a_1t}  
L_{\alpha} ( (x-x_0) (gt)^{-1/\alpha} ) +O(\varepsilon^2) \Bigr) ,
\ee
and the asymptotic is
\be
Z(x,t)=e^{i(a_2t+\theta(0))} \Bigl( 
a^{1/2}_1 + \varepsilon gt e^{-2a_1t} \pi^{-1} \Gamma(1+\alpha) (x-x_0)^{-\alpha-1} 
+O(\varepsilon^2) \Bigr) , \quad x \rightarrow \infty .
\ee
This solution shows that the long-wave modes approach to
the limit cycle exponentially with time.
For $t=1/(2a_1)$, we have the maximum of $|Z(x,t)|$ with respect to time: 
\be
\max_{t>0} |Z(x,t)|=
a^{1/2}_1 + \varepsilon g  \frac{\Gamma(1+\alpha)}{2\pi e} (x-x_0)^{-\alpha-1} 
+O(\varepsilon^2) .
\ee
As the result, we have the power law decay with respect to coordinate
for the space structures near the limit cycle $|Z|=a^{1/2}_1 $.

\begin{figure}
\rotatebox{270}{\includegraphics[width=15 cm,height=15 cm]{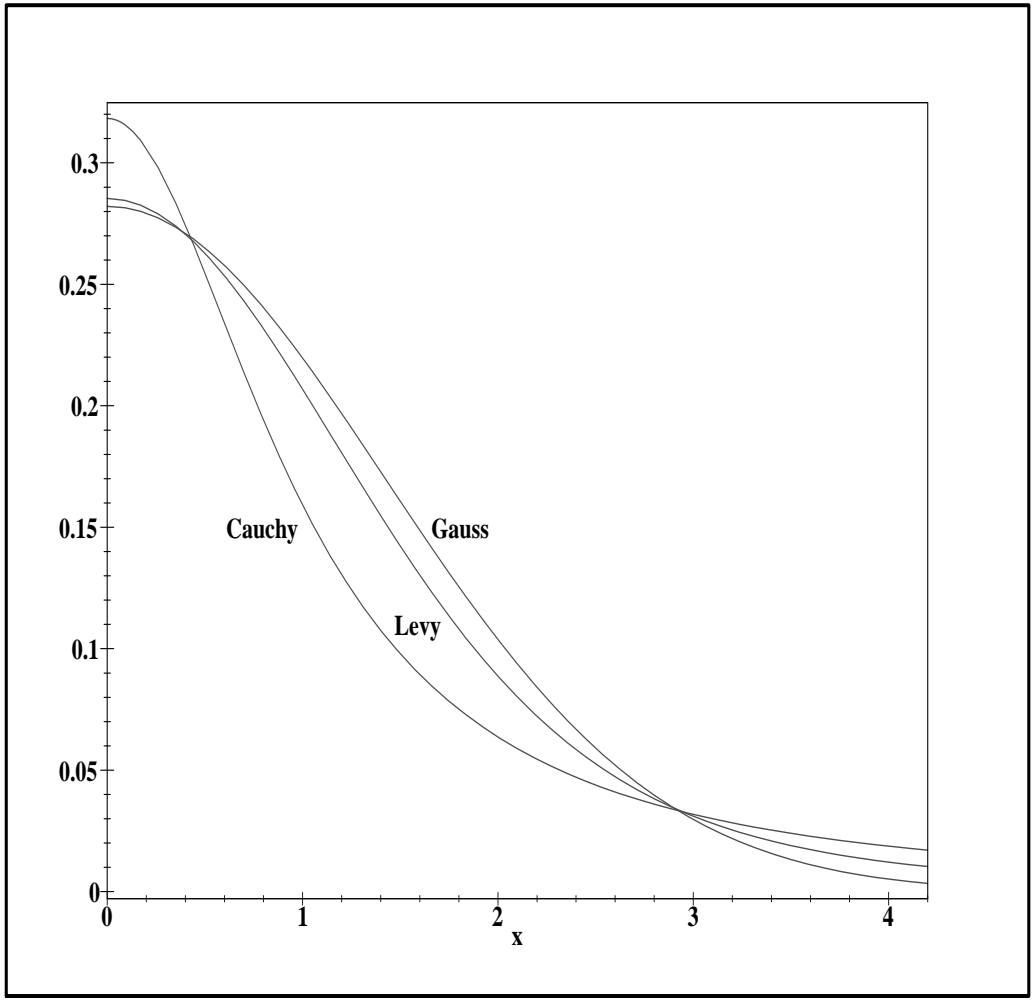}}
\caption{\label{fig1} Gauss p.d.f. ($\alpha=2$), 
Levy p.d.f. ($\alpha=1.6$), and Cauchy p.d.f. ($\alpha=1.0$).  
Levy for $\alpha=1.6$ lies between Cauchy and Gauss p.d.f.
In the asymptotic $x \rightarrow \infty$ and $x>3$ on the plot, 
the upper curve is Cauchy p.d.f, the lower curve is Gauss p.d.f.
}
\end{figure}

\section{Nonlinear long-range interaction and fractional phase equation}

In this section, we would like to show one more
application of the replacement of dynamical equation by the 
fractional ones for a chain with long-range interaction.
The model was first considered in \cite{Win,Kur1,Kur2} 
with application in biology and chemistry.
This model has additional interest since it can be reduced 
to a chain of interacting spins.
In this section, we will derive the fractional phase (spin) equations
and leave their solution for the future publication.

\subsection{Nonlinear nonlocal phase coupling}

Let us consider the phase equation
\be \label{F1}
\frac{d}{dt}\theta_n(t)=
\omega_n+g\sum_{m \not=n} J_{\alpha}(n-m) \sin(\theta_n-\theta_m) ,
\ee
where $\theta_n$ denotes the phase of the $n$-th oscillator, 
$\omega_n$ its natural frequency, and
\be
J_{\alpha}(n)=|n|^{-\alpha-1} .
\ee
For $\alpha=-1$, Eq. (\ref{F1}) defines the Kuramoto model \cite{Kur1,Kur2,Str1}
with sinusoidal non-local coupling (infinite radius of interaction). 
We can rewrite Eq. (\ref{F1}) for classical spin-like variables
\be \label{F4}
s_n(t)=e^{i\theta_n(t)}. 
\ee
Using 
\be \label{F5}
s^{-1}_n=s^*_n, \quad \theta_n=-i \ln s_n ,
\ee
we get
\be \label{F6}
\frac{d}{dt} \theta_n=
-i \frac{d}{dt} \ln s_n =-i s^{-1}_n \frac{d}{dt} s_n =
-is^*_n \frac{d}{dt} s_n ,
\ee
and
\be \label{F7}
\sin(\theta_n-\theta_m) =\frac{1}{2i} 
\left( e^{i(\theta_n-\theta_m)}+ e^{-i(\theta_n-\theta_m)}  \right)=
\frac{1}{2i} \left( s_n s^*_m+ s^*_n s_m  \right) .
\ee
Then Eq. (\ref{F1}) is
\be \label{F8}
s^*_n \frac{d}{dt} s_n=i\omega_n +
\frac{g}{2} \sum_{m\not=n} 
\frac{1}{|n-m|^{\alpha+1}} [ s_n s^*_m+ s^*_n s_m  ] .
\ee
This equation describes the long-range interaction of spin variables.
We also will call Eq. (\ref{F8}) as the phase coupling 
equation since $|s_n|^2=const$.
Thermodynamics of the model of classical spins with long-range
interactions (see for discussion Appendix 1) has been studied more than thirty years.
An infinite one-dimensional Ising model with long-range interactions 
was considered by Dyson \cite{Dyson}.
The $d$-dimensional classical Heisenberg model with long-range 
interaction is  described in Refs. \cite{J,CMP}, and
its quantum generalization with long-range interaction decreasing  as 
$|n|^{-\alpha}$ can be found in \cite{NakTak}.

\subsection{Phase-coupled oscillatory medium 
with nonlinear long-range interaction}

Let us derive an equation for continuous medium 
that consists of oscillators of (\ref{F1}) or (\ref{F8}) 
type with nonlinear long-range interaction. 
The medium can be defined by the field 
\be
S(x,t)=\frac{1}{2\pi} \int^{+\infty}_{-\infty} dk \ 
e^{ikx} \sum^{+\infty}_{n=-\infty} e^{-ikn} s_n(t) .
\ee
We also will need the following momentum representations
\be \label{w} 
a(k,t)=\sum^{\infty}_{n=-\infty} e^{-ikn} s_n(t) . \ee
For left hand side of (\ref{F8}), we get (see Appendix 2):
\be \label{K5s}
\frac{1}{2\pi} \int^{\infty}_{-\infty} dk e^{ikx}
\sum^{\infty}_{n=-\infty} e^{-ikn} s^{*}_n \frac{d}{dt}s_n= 
S^*(x,t) \frac{d}{dt} S(x,t) .
\ee
For the interaction term, we obtain (Appendix 2):
\[ 
\frac{1}{2\pi} \int^{\infty}_{-\infty} dk e^{ikx}
\sum^{\infty}_{n=-\infty} e^{-ikn} 
\sum_{m\not=n} \frac{1}{|n-m|^{\alpha+1}} s^*_n s_m  =
S^*(x,t) \frac{1}{2\pi} \int^{\infty}_{-\infty} dk_1 
a(k_1,t) \tilde J_{\alpha}(k_1) e^{ik_1x} =
\]
\be \label{K9b}
=S^*(x,t) \left( 2 \zeta(\alpha+1)S(x,t) -
a_{\alpha} \frac{\partial^{\alpha}}{\partial |x|^{\alpha}} S(x,t)+
2\sum^{\infty}_{n=0} \frac{\zeta(\alpha+1-2n)}{(2n)!} 
\frac{\partial^{2n}}{\partial x^{2n}} S(x,t) \right) ,
\ee
where we use (\ref{D2}) for
$\tilde J_{\alpha}(k)$ which is defined by (\ref{C5}), and
\be \label{aalp}
a_{\alpha}=2\Gamma(-\alpha)\cos(\pi \alpha/2) . 
\ee

For the term $\omega_n$, we use 
\be
\omega(x)=\frac{1}{2\pi} \int^{+\infty}_{-\infty} dk e^{ikx}  
\sum^{\infty}_{n=-\infty} e^{-ikn} \omega_n . 
\ee
If all oscillators have the same natural frequency $\omega_n=\omega$,
then $\omega (x)=\omega$.

As the result,
\[ S^{*}(x,t)\frac{\partial}{\partial t} S(x,t)=
i\omega(x)-f_{\alpha} S^{*}(x,t) S(x,t)- \]
\[ -g_{\alpha} 
\left(S^*(x,t) \frac{\partial^{\alpha}}{\partial |x|^{\alpha}} S(x,t)
+S(x,t) \frac{\partial^{\alpha}}{\partial |x|^{\alpha}} S^*(x,t)
\right)+ \]
\be \label{P1}
+g\sum^{\infty}_{n=1} \frac{\zeta(\alpha+1-2n)}{(2n)!} 
\left(S^*(x,t) \frac{\partial^{2n}}{\partial |x|^{2n}} S(x,t)
+S(x,t) \frac{\partial^{2n}}{\partial |x|^{2n}} S^*(x,t)
\right) ,
\ee
where
\be \label{P2}
f_{\alpha}=2g\zeta(\alpha+1), \quad
g_{\alpha}=(1/2) a_{\alpha} g=g \Gamma(-\alpha) \cos(\pi\alpha/2) .
\ee
Eq. (\ref{P1}) is a fractional equation for 
oscillatory medium with long-range interacting spins (\ref{F8}).
We can call (\ref{P1}) the fractional phase equation.

In the infrared approximation ($k \rightarrow 0$), we can use (\ref{D2}):
\be \label{D2n}
\tilde J_{\alpha}(k)\approx 2\Gamma(-\alpha)\cos(\pi \alpha/2) |k|^{\alpha}+
2 \zeta(\alpha+1) ,  \quad 0< \alpha <2 , \quad \alpha \not=1 ,
\ee
and equation (\ref{P1}) is reduced to
\be\label{P1i}
S^{*}(x,t)\frac{\partial}{\partial t} S(x,t)=
i\omega(x)-f_{\alpha} 
- g_{\alpha} \left(S^*(x,t) \frac{\partial^{\alpha}}{\partial |x|^{\alpha}} S(x,t)
+S(x,t) \frac{\partial^{\alpha}}{\partial |x|^{\alpha}} S^*(x,t) \right) , \ee
where $0< \alpha< 2$, $\alpha \not=1$.

\section{Conclusion}

One-dimensional chain of interacting objects, say oscillators, 
can be considered as a benchmark for numerous applications 
in physics, chemistry, biology, etc.
All considered models were related mainly to the oscillating objects
with long-range power-wise interaction, i.e., with
forces proportional to $1/|n-m|^s$ and $2<s<3$. 
A remarkable feature of this interaction is the existence
of a transform operator that replaces the set of 
coupled individual oscillator equations 
into the continuous medium equation
with fractional space derivative of 
order $\alpha=s-1$, where $0<\alpha<2$, $\alpha\not=1$.
Such transformation is an approximation and 
it appears in the infrared limit for wave number $k \rightarrow 0$.
This limit helps to consider different models and related phenomena
in a unified way applying
different tools of fractional calculus.

A nontrivial example of general property of fractional linear equation 
is its solution with a power-wise decay along the space coordinate.
From the physical point of view that means a new type of 
space structures or coherent structures.
The scheme of equations with fractional derivatives
includes either effect of synchronization \cite{Pik1},
breathers \cite{Br3,Br4}, fractional kinetics \cite{Zaslavsky1},
and others.

Discrete breathers are periodic space-localized oscillations 
that arise in discrete and continuous nonlinear systems. 
Their existence was proven in Ref. \cite{Br1}. 
Discrete breathers have been widely studied in systems 
with short-range interactions (for a review, see \cite{Br2,Br3}). 
Energy and decay properties of discrete breathers in systems with long-range 
interactions have also been studied in the framework of the Klein-Gordon 
\cite{Br4,Br5}, and the discrete nonlinear Schrodinger equations \cite{Br6}.
Therefore, it is interesting to consider breathers solution 
in systems with long-range interactions in infrared approximation.

We also assume that the obtained transform operator can be 
used for improvement of different scheme of simulations 
for equations with fractional derivatives.

\section*{Acknowledgments}

We are thankful to N. Laskin for  useful discussions and comments.
This work was supported by the Office of Naval Research,
Grant No. N00014-02-1-0056, the U.S. Department
of Energy Grant No. DE-FG02-92ER54184, and the NSF
Grant No. DMS-0417800. 
V.E.T. thanks the Courant Institute of Mathematical Sciences
for support and kind hospitality.


\section*{Appendix 1: Spin chain models with long-range interactions}

The model of classical spins with long-range
interactions has been studied for more than thirty years.
An infinite one-dimensional Ising model 
with long-range interactions was considered by Dyson \cite{Dyson}.
Existence of a phase-transition was proved for 
an infinite linear chain of spins $s_n$ 
with an interaction Hamiltonian
\be \label{H4A}
H_{int} = - \sum_{n,\ m \not=n} J_{\alpha}(n-m) s_n s_m , \quad 
J_{\alpha}(n)=|n|^{-\alpha-1} . 
\ee
It was proved that the Ising model (\ref{H4A}) has a phase
transition of the first order when $0 < \alpha < 1$, i.e. for fairly
distant interactions.

For the $d$-dimensional ($d=1,2$) classical Heisenberg
model with long-range interaction 
\be \label{XYmodel}
H_{int}=-\sum_{n , \ m\not=n} J_{\alpha}(n-m) \ {\bf s}_n {\bf s}_m , 
\ee
the existence of the first order phase transition for $d<\alpha<2d$
and nonexistence for $\alpha\ge 2d$ was proved in Refs. \cite{J,CMP}.

The quantum XY-model with long-range interaction (\ref{XYmodel})
was considered in Refs. \cite{NakTak}.
This model in the limit $\alpha \rightarrow \infty$ 
corresponds to nearest-neighbor interaction.
The spin operators ${\bf s}_n$ at each lattice site
can be replaced by boson creation and annihilation operators:
\[ s^x_n+is^y_n=\sqrt{2}a^+_n, \quad s^x_n-is^y_n=\sqrt{2}a_n .\]
The site representation can be replaced by
the momentum representation using Fourier transformation
\[ a^+_n=\frac{1}{\sqrt{N}} \sum_k e^{-ikn} a^{+}(k) ,\quad 
a_n=\frac{1}{\sqrt{N}} \sum_k e^{ikn} a(k) , \]
where $a^+(k)$, $a(k)$ are the corresponding Fourier transforms
of operators $a^+_n$, $a_n$, and $N$ is a number of particles.
Then the interaction part of the Hamiltonian can be rewritten as
\be \label{H6A}
H_{int}=\sum_k E(k) a^+(k) a(k)
\ee
with the dispersion relation 
\be \label{H7A}
E(k)=[\tilde J_{\alpha}(0)- \tilde J_{\alpha}(k)] , \quad 
\tilde J_{\alpha}(k)=2\sum^N_{n=1} J_{\alpha}(n) \cos(kn) .
\ee 
In the thermodynamic limit $N\rightarrow \infty$
the dispersion relation (\ref{H7A}) becomes
\be \label{H8A}
E(k)=2g \sum^{\infty}_{n=1} \frac{1-\cos(kn)}{n^{\alpha+1}} .
\ee
For small $k$ (infrared limit $|k| \rightarrow 0$) and $0<\alpha<2$, 
$\alpha \not=1$, we have
\be \label{H9A}
E(k)\approx j_{\alpha} |k|^{\alpha}, 
\ee 
where
\be
j_{\alpha}=\frac{\pi g}{\Gamma(\alpha) \cos(\pi \alpha/2)} .
\ee
For oscillatory medium, Eq. (\ref{H9A}) gives the fractional
Riesz derivatives \cite{Lask} of order $\alpha$ 
(compare to (\ref{D10})).

\section*{Appendix 2: Derivation of fractional phase equation}

The left hand side of (\ref{F8}) is
\be \label{K1}
s^{*}_n \frac{d}{dt}s_n=\frac{1}{(2\pi)^2} \int^{+\infty}_{-\infty} 
dk_1 dk_2 e^{-ink_1} e^{ink_2}
a^*(k_1,t) \frac{d}{dt} a(k_2,t) ,
\ee
or for its Fourier transform
\be \label{K2}
\sum^{\infty}_{n=-\infty} e^{-ikn} s^{*}_n \frac{d}{dt}s_n=
\frac{1}{(2\pi)^2} \int^{\infty}_{-\infty} dk_1 dk_2 
\sum^{\infty}_{n=-\infty} e^{in(k_2-k_1-k)}
a^*(k_1,t) \frac{d}{dt} a(k_2,t) =
\ee
\[
=\frac{1}{2\pi} \int^{\infty}_{-\infty} 
dk_1 dk_2 \delta(k_2-k_1-k)a^*(k_1,t) \frac{d}{dt} a(k_2,t) = 
\]
\be \label{K4}
=\frac{1}{2\pi} \int^{\infty}_{-\infty} dk_2 a^*(k_2-k,t) \frac{d}{dt} a(k_2,t) .
\ee
Then
\[
\frac{1}{2\pi} \int^{\infty}_{-\infty} dk e^{ikx}
\sum^{\infty}_{n=-\infty} e^{-ikn} s^{*}_n \frac{d}{dt}s_n= 
\frac{1}{2\pi} \int^{\infty}_{-\infty} dk e^{ikx}
\frac{1}{2\pi} \int^{\infty}_{-\infty} dk_2 a^*(k_2-k,t) \frac{d}{dt} a(k_2,t) =
\]
\[
=\frac{1}{2\pi} \int^{\infty}_{-\infty} dk_2 \frac{d}{dt} a(k_2,t) 
\frac{1}{2\pi} \int^{\infty}_{-\infty} dk a^*(k_2-k,t) e^{ikx}=
\]
\be \label{K5}
=\frac{1}{2\pi} \int^{\infty}_{-\infty} dk_2 e^{ik_2x} \frac{d}{dt} a(k_2,t) 
\frac{1}{2\pi} \int^{\infty}_{-\infty} dk_1 a^*(k_1,t) e^{-ik_1 x}=
S^*(x,t) \frac{d}{dt} S(x,t) ,
\ee
where we use 
\be
S(x,t)=\frac{1}{2\pi} \int^{+\infty}_{-\infty} e^{ikx} a(k,t) dk .
\ee

Let us consider the interaction terms 
\be \label{K6}
\sum_{m\not=n} \frac{1}{|n-m|^{\alpha+1}} s_m(t)s^*_n(t)=
\sum_{m\not=n} \frac{1}{|n-m|^{\alpha+1}} 
\frac{1}{(2\pi)^2} \int^{\infty}_{-\infty} dk_1 dk_2 e^{imk_1} e^{-ink_2}
a(k_1,t) a^*(k_2,t) .
\ee
Multiplying this equation by $\exp(-ikn)$, 
and summing over $n$ from $-\infty$ to $+\infty$, we obtain
\[
\sum^{\infty}_{n=-\infty} e^{-ikn} 
\sum_{m\not=n} \frac{1}{|n-m|^{\alpha+1}} s^*_n s_m  =
\]
\[
=\sum^{\infty}_{n=-\infty} e^{-ikn} \sum_{m\not=n} \frac{1}{|n-m|^{\alpha+1}} 
\frac{1}{(2\pi)^2} \int^{\infty}_{-\infty} dk_1 dk_2 e^{imk_1} e^{-ink_2}
a(k_1,t) a^*(k_2,t) =
\]
\[
=\frac{1}{(2\pi)^2} \int^{\infty}_{-\infty} dk_1 dk_2 
a(k_1,t) a^*(k_2,t) 
\sum^{\infty}_{n=-\infty} \sum_{m\not=n} \frac{1}{|n-m|^{\alpha+1}} 
e^{imk_1} e^{-in(k+k_2)}=
\]

\[
=\frac{1}{(2\pi)^2} \int^{\infty}_{-\infty} dk_1 dk_2 
a(k_1,t) a^*(k_2,t) 
\sum^{\infty}_{n=-\infty}e^{-in(k+k_2)} \sum_{m\not=n} \frac{1}{|n-m|^{\alpha+1}} 
e^{imk_1} =
\]

\[
=\frac{1}{(2\pi)^2} \int^{\infty}_{-\infty} dk_1 dk_2 
a(k_1,t) a^*(k_2,t) 
\sum^{\infty}_{n=-\infty}e^{-in(k+k_2-k_1)} 
\sum_{m^{\prime}\not=0} \frac{1}{|m^{\prime}|^{\alpha+1}} 
e^{im^{\prime}k_1} =
\]

\[
=\frac{1}{(2\pi)^2} \int^{\infty}_{-\infty} dk_1 dk_2 
a(k_1,t) a^*(k_2,t) \sum^{\infty}_{n=-\infty}e^{-in(k+k_2-k_1)} \tilde J_{\alpha}(k_1) =
\]
\[
=\frac{1}{2\pi} \int^{\infty}_{-\infty} dk_1 dk_2 
a(k_1,t) a^*(k_2,t) \delta(k+k_2-k_1) \tilde J_{\alpha}(k_1) =
\]
\be \label{K7}
=\frac{1}{2\pi} \int^{\infty}_{-\infty} dk_1 
a(k_1,t) a^*(k_1-k,t)   \tilde J_{\alpha}(k_1) ,
\ee
where $\tilde J_{\alpha}(k)$ is defined by (\ref{C5}).
Multiplying the final expression (\ref{K7}) by $\exp(ikx)$, 
and integrating over $k$ from $-\infty$ to $\infty$, 
we get finally
\[
\frac{1}{2\pi} \int^{\infty}_{-\infty} dk e^{ikx}
\frac{1}{2\pi} \int^{\infty}_{-\infty} dk_1 
a(k_1,t) a^*(k_1-k,t)  \tilde J_{\alpha}(k_1) =
\]
\[
=\frac{1}{2\pi} \int^{\infty}_{-\infty} dk_1 
a(k_1,t) \tilde J_{\alpha}(k_1) 
\frac{1}{2\pi} \int^{\infty}_{-\infty} dk e^{ikx} a^*(k_1-k,t)  = \]
\[
=\frac{1}{2\pi} \int^{\infty}_{-\infty} dk_1 
a(k_1,t) \tilde J_{\alpha}(k_1) e^{ik_1x}
\frac{1}{2\pi} \int^{\infty}_{-\infty} dk e^{-ik^{\prime}x} a^*(k^{\prime},t) =
\]

\[ 
=S^*(x,t) \frac{1}{2\pi} \int^{\infty}_{-\infty} dk_1 
a(k_1,t) \tilde J_{\alpha}(k_1) e^{ik_1x} =
\]
\be \label{K9}
=S^*(x,t) \left( 2 \zeta(\alpha+1)S(x,t) -
a_{\alpha} \frac{\partial^{\alpha}}{\partial |x|^{\alpha}} S(x,t)+
2\sum^{\infty}_{n=0} \frac{\zeta(\alpha+1-2n)}{(2n)!} 
\frac{\partial^{2n}}{\partial x^{2n}} S(x,t) \right) ,
\ee
where $a_{\alpha}$ is defined by (\ref{aalp}), and we use (\ref{D2}). 
This expression is just the equation (\ref{K9b}).

\end{document}